\documentclass[%
reprint,
10pt,aps,showpacs, %Me
nofootinbib,
 amsmath,amssymb,
 aps,
amsfonts, axodraw,graphicx, %Me
prd, %Me
epsf, %Me
floats, %Me
floatfix,
]{revtex4-2}

\usepackage{graphicx}% Include figure files
\usepackage{dcolumn}% Align table columns on decimal point
\usepackage{bm}% bold math
\usepackage{hyperref} %---Me (To jump into citation, formula, section and etc)
\usepackage{lipsum} %-----Footnote not to break
\usepackage{xcolor} %-----Colorful text
\DeclareMathOperator{\Tr}{Tr}
\DeclareMathOperator{\tr}{tr}
\renewcommand{\theequation}{\thesection.\arabic{equation}}
\newcounter{saveeqn}
\newcommand{\add}{\addtocounter{equation}{1}}
\newcommand{\alphaeqn}{\setcounter{saveeqn}{\value{equation}}%
\setcounter{equation}{0}%
\renewcommand{\theequation}{\mbox{\thesection.\arabic{saveeqn}{\alpha{equation}}}}}
\newcommand{\reseteqn}{\setcounter{equation}{\value{saveeqn}}%
\renewcommand{\theequation}{\thesection.\arabic{equation}}}

\begin{document}

\preprint{APS/123-QED}

\title{Chiral symmetry breaking and phase diagram of dual chiral density wave in a rotating quark matter}%rotating system}% Force line breaks with \\
%\thanks{A footnote to the article title}%

\author{S.~M.~A.~Tabatabaee Mehr}\email{tabatabaee@ipm.ir}
\affiliation{School of Particles and Accelerators, Institute for Research in Fundamental Sciences (IPM), P.O. Box 19395-5531, Tehran, Iran}

\date{\today}% It is always \today, today,
             %  but any date may be explicitly specified

\begin{abstract}
We study the inhomogeneous phase of a two-flavor quark matter under rotation at finite temperature and density using the Nambu-Jona-Lasinio model. To do this, we consider the chiral broken phase, in particular, described by the so-called dual chiral density wave which is formed as a standing wave of simultaneous scalar and pseudoscalar condensates. The solution of the corresponding Dirac equation as well as the energy spectrum found in the mean-field approximation. We then use the thermodynamic potential calculated for this model, to study the $\mu$ and $\Omega$ dependence of constituent mass and the wave vector at $T = 0$. We find there exist two islands in the $\mu - \Omega$ plane that the dual-chiral density wave survives. The first region lies at intermediate densities and small $\Omega$. We observe, by increasing the angular velocity of matter, dual-chiral density wave forms in regions with smaller chemical potential. On the other hand, in contrast to the former, the second region is located at the large $\Omega$ and small densities. Finally, we study this phase of quark matter at finite temperature and present $T-\mu$, $T-\Omega$, and $\mu-\Omega$ phase portraits of a hot-rotating quark matter at finite density.
\end{abstract}

%\keywords{Suggested keywords}%Use showkeys class option if keyword
                              %display desired
\maketitle

%\tableofcontents

\section{INTRODUCTION}
%%%%
\par
Mapping the phase structure of QCD at finite temperature and nonzero chemical potential in a rotating system is one of the recent challenges in the condensed matter of strong interaction. The main interest arises from the relevant experiments on the global spin polarization of produced particles \cite{STAR:2017ckg}. According to these measurements, in noncentral heavy-ion collision (HIC) at the Relativistic Heavy Ion Collider (RHIC) the colliding nuclei deposit a large angular momentum on the order of $L \approx 10^{3} \hbar$ in the produced quark-gluon plasma (QGP). As a consequence of nonvanishing angular momentum in the plasma of quarks and gluons, the shear flow of longitudinal momentum arises which in turn, leads to the generation of of strong vortical structure $\omega \approx 10^{22} s^{-1}$. Such a large vorticity may have some nontrivial effects on the properties of quark matter. Possible effects of rotation include the chiral vortical effect \cite{Vilenkin:1979ui,Vilenkin:1980zv,Fukushima:2018grm}, chiral vortical wave \cite{Jiang:2015cva}, the spin polarization of particles \cite{Becattini:2020ngo,Huang:2020dtn}, pion condensate \cite{Liu:2017spl,Cao:2019ctl}, the confinement-deconfinement transition \cite{Chernodub:2020qah,Braguta:2021jgn,Chen:2022smf}, and its influence on the scalar chiral condensate of quark and antiquark and consequently the reduction in the transition temperature of chiral symmetry restoration \cite{Jiang:2016wvv,Chernodub:2016kxh,Wang:2018sur,Zhang:2020hha,Sadooghi:2021upd,Mehr:2022tfq}. It is, therefore, the purpose of this paper to scrutinize the effect of rotation on the chiral condensate, in particular at moderate densities, and study the phases of quark matter in this region of phase space.
\par
There are pieces of evidence in theoretical studies of the thermodynamic properties of systems under rigid rotation, which imply the suppression of scalar condensate \cite{Jiang:2016wvv,Sadooghi:2021upd}. This phenomenon is related to the alignment of spin of particles in the direction of angular velocity which is known as the Barnett effect \cite{Barnett:1915}. Therefore, the rotation tends to decrease the transition temperature of the chiral symmetry-broken ($\chi SB$) phase to the chiral symmetry restored ($\chi SR$) phase. Moreover, as shown in \cite{Wang:2018sur}, in the $T - \mu$ plane of the phase diagram, with increasing the angular velocity of the system the location of the critical point (CP) is shifted to the region of phase space with fixed chemical potential but smaller temperature. Hence, in addition to the reduction in the phase boundary due to the decrease in the transition temperature, the change in the location of CP manifests itself, in particular, in the broadening of the second-order phase boundary. Apart from the nature of the phase structure of quark matter under rotation, another feature of this system is the breakdown of translational invariance in a plane orthogonal to the angular velocity whose signature manifests as a radial inhomogeneity in the constituent mass \cite{Zhang:2020hha,Sadooghi:2021upd,Chen:2022mhf}.
\par
Apart from the radial inhomogeneity caused by rotation, in a parallel development, the possibility of the crystalline phase of quark matter, which is described by an inhomogeneous-order parameter, at intermediate chemical potential is the subject of intensive studies (for a review see \cite{Ferrer:2021mpq,Buballa:2014tba} and the references therein). Conventionally, the chiral condensate, which is the pair of left (right)-handed quarks and the right (left)-handed antiquarks, is assumed to be uniform and homogeneous in space. At finite density $\mu$, however, the pairing of the quark and antiquark cost energy is of $2 \mu$. Thus, by increasing the chemical potential the energy cost is no longer compensated by the condensation energy which at some critical chemical potential the homogeneous chiral condensate is not the favored ground state of quark matter. Departure from this line of work led us to the spatially modulated condensate in the phase diagram of QCD. At finite chemical potential, in particular, another variant of condensate between the quark with momentum $| \vec p | \sim \mu$ and quark hole with the same momentum $\vec p$ near the Fermi surface with finite total momentum $P \sim 2 \mu$ arises \cite{Tatsumi:2004dx}. This particular form of condensate, which appears only at finite density, varies in space as $ \exp ( 2 i \vec p \cdot \vec x)$ and is dubbed as chiral density wave. The dual chiral density wave (DCDW), one particular type of density wave, appears as a simultaneous spatially varying scalar and pseudoscalar condensate. This form of modulation, particularly described by $M(\textbf{x}) = -G \left(\langle \bar \psi \psi \rangle+ i \gamma^{5} \langle \bar \psi i \gamma^{5} \psi \rangle \right) \equiv -G \Delta (\textbf{x}) e^{i \gamma^{5} \vartheta (\textbf{x})} $, transforms the scalar (pseudoscalar) condensate to the pseudoscalar (scalar) one and stays in a hypersphere of constant radius $\Delta$ in chiral space \cite{Broniowski:1990dy,Nakano:2004cd,Carlomagno:2015nsa,Yoshiike:2017kbx,Moreira:2013ura}. Stability analysis of the DCDW against the thermal fluctuations shows that this phase of quark matter, in particular, with a one-dimensional (1d) modulation exhibit Landau-Peierls instability. This arises as a consequence of soft-transverse fluctuation modes in the spectrum, which in turn wipes out the long-range order \cite{Lee:2015bva}. However, the existence of Landau-Peierls instability does not rule out the 1d inhomogeneous phase. It is argued, in analogy with the smectic liquid crystals, that this form of modulation may exist and, therefore, realized in a quite different form with a quasilong-range order \cite{Lee:2015bva,Baym:1982ca}. Moreover, it is shown that in the presence of a uniform magnetic field the modulation along the magnetic field is favored and the magnetic dual-chiral density wave (MDCDW) phase is stabilized \cite{Frolov:2010wn,Gyory:2022hnv}. According to the stability analysis of the DCDW in a uniform magnetic field, it turns out that this phase demonstrates the absence of the Landau-Peierls instability arising from the lack of soft-transverse fluctuation modes \cite{Ferrer:2019zfp}.
\par
Using various QCD inspired models, i.e. Gross-Neveu (GN) \cite{Thies:2006ti,Basar:2008im,Buballa:2020nsi}, the Nambu-Jona-Lasinio (NJL) model \cite{Tatsumi:2004dx,Nakano:2004cd,Karasawa:2013zsa,Buballa:2015awa,Carignano:2019ivp}, the quark-meson (QM) model \cite{Broniowski:1990dy,Carignano:2014jla} as well as the Schwinger-Dyson approach \cite{Muller:2013tya} and considering different forms of inhomogeneity, it is argued at moderate densities and low temperature, an island of spatially modulated constituent mass is a favorable ground state of QCD. Thus, this region in the phase portrait of quark matter exhibits an inhomogeneous phase of chiral condensate. 
\par
It is thus the purpose of this paper to fill the gap and find a connection between these two approaches to study the possibility of an inhomogeneous phase of quark matter in a rotating system. Therefore, in this paper, we proceed further to study the effect of rotation on the inhomogeneous phase of quark matter, in particular DCDW, and investigate the interplay between the chemical potential and angular velocity at zero and finite temperature. To do this, we utilize a two-flavor NJL model defined in a rotating system. Assuming the system under consideration develops an inhomogeneous chiral condensate according to the DCDW, we numerically determine the phase portrait of the inhomogeneous phase of quark matter.
\par
The organization of the paper is as follows: We devote Sec. \ref{sec2A} to introduce a two-flavor Nambu-Jona-Lasinio (NJL) model in a rotating system. Moreover, in this section, after providing the general concepts on our framework and fixing our notation in analogy to \cite{Sadooghi:2021upd}, the thermodynamic potential of our model is determined with formation of dual chiral density wave in the mean field approximation. In Sec. \ref{sec2B}, the solution of Dirac equation in presence of DCDW is discussed. In Sec. \ref{sec3A}, after fixing the free parameters of the model, coupling constant $G$ and the momentum cutoff $\Lambda$ introduced in proper time regularization, we numerically find the global minimum of thermodynamic potential presented in Sec. \ref{sec2A} for a cold quark matter. We then present the numerical results for chemical potential dependence of constituent mass together with the wave vector for different values of angular velocity. Moreover, we discuss the possibility to form DCDW in smaller values of chemical potentials in a fast rotating matter. As a by-product, the phase diagram of DCDW in the $\mu - \Omega$ plane is presented in this section. Then, Sec. \ref{sec3B} is devoted to the study of the phase diagram of DCDW at finite T, $\mu$ as well as $\Omega$. In Sec. \ref{sec4} we summarize our main results and discuss some concluding remarks.
%%%%%%%%%%%%%
\section{Two flavor dual-chiral density wave in a rotating quark matter}
\setcounter{equation}{0}
%%%%%%%%%%%%%
%%%%%%%%%%%%%
\subsection{The model} \label{sec2A}
%%%%%%%%%%%%%
It is believed at moderate densities, there is a possibility of an inhomogeneous chiral condensate. In the present paper, we focus on the effect of rotation on the inhomogeneous phase of quark matter using the two-flavor  NJL model.  For this purpose, we study the quark matter in the corotating frame which is assumed to be rotating uniformly with constant angular velocity $\Omega$ about the $z$-direction. Performing the coordinate transformation of $\varphi = \varphi_{M} - \Omega \, t$ to the Minkowski line element, the corresponding corotating line element is given by
\begin{eqnarray}\label{A1}
ds^2&=&g_{\mu\nu}dx^{\mu}dx^{\nu}=\left(1-r^{2}\Omega^{2}\right)dt^2-dx^2\nonumber\\
&&+2\Omega ydtdx-dy^2-2\Omega xdtdy-dz^{2},
\end{eqnarray}
with  $x^{\mu}=(t,x,y,z)=(t,r\cos\varphi,r\sin\varphi,z)$, where $r$ and $\varphi$ are the cylindrical variables. %The world-line in \eqref{A1} is equivalent to the metric
%\begin{eqnarray}\label{A2}
%\hspace{-0.5cm}g_{\mu\nu}=\left(
%\begin{array}{cccc}
%1-\Omega^2(x^2+y^2)&+\Omega y&-\Omega x&0\\
%+\Omega y&-1&0&0\\
%-\Omega x&0&-1&0\\
%0&0&0&-1
%\end{array}
%\right).
%\end{eqnarray}
Using the above metric together with the concepts and notations developed in the curved space, the Lagrangian density of a two-flavor NJL model at finite density is given by
\begin{eqnarray}\label{A3}
\mathcal{L} = \bar \psi \left[ i \gamma^{\mu} D_{\mu} + \mu \gamma^{0}  \right] \psi + \frac{G}{2} \left[ \left( \bar \psi \psi  \right)^{2} + \left( \bar \psi  i\gamma^{5} \bm{\tau} \psi  \right)^{2}  \right], \nonumber \\
\end{eqnarray}
where $\mu$ is the chemical potential and $\bm{\tau} = (\tau_{1},\tau_{2},\tau_{3})$ are the Pauli matrices in isospin space. Moreover, the covariant derivative is defined as $D_{\mu} = \partial_{\mu} + \Gamma_{\mu} $. The affine connection $\Gamma_{\mu}$ appearing in the covariant derivative is given in terms of the spin connection $\omega_{\mu ab}$ and vierbeins $e^{\mu}_{~a}$ as
\begin{eqnarray}\label{A4}
\Gamma_{\mu}&\equiv&-\frac{i}{4}\omega_{\mu ab}\sigma^{ab}, \, \omega_{\mu ab}\equiv g_{\alpha\beta}e^{\alpha}_{~a}\left(\partial_{\mu}e^{\beta}_{~b}+\Gamma^{\beta}_{\mu\nu}e^{\nu}_{~b}\right). \nonumber \\
\end{eqnarray}
Here, spin matrices defined as $\sigma^{ab}\equiv\frac{i}{2}[\gamma^{a},\gamma^{b}]$ and the Christoffel connection given by $\Gamma^{\beta}_{\mu\nu}\equiv \frac{1}{2}g^{\beta\sigma}\left(\partial_{\mu}g_{\sigma\nu}+\partial_{\nu}g_{\mu\sigma}-\partial_{\sigma}g_{\mu\nu}\right)$. Let us note that, the Greek indices $\mu=t,x,y,z$ and the Latin indices $a=0,1,2,3$ appearing in this notation refer to the general and the Cartesian coordinate in the tangent space, respectively. As it turns out, the components of vierbein satisfy the identity $\eta_{ab}=g_{\mu\nu}e^{\mu}_{~a}e^{\nu}_{~b}$ with the Minkowski metric $\eta_{ab}=\mbox{diag}\left(1,-1,-1,-1\right)$. Moreover, they connect the general coordinate to the Cartesian one in the tangent space by  $x^{\mu}=e^{\mu}_{~a}x^{a}$. Adopting the vierbeins in Cartesian gauge %\cite{Sadooghi:2021upd}
\begin{eqnarray}\label{A5}
&&  e^{t}_{~0}=e^{x}_{~1}=e^{y}_{~2}=e^{z}_{~3}=1, \nonumber\\
&& e^{x}_{~0}= \Omega y,\quad e^{y}_{~0}=- \Omega x,
\end{eqnarray}
together with the nonvanishing components of $\Gamma^{\mu}_{\alpha \beta}$,
\begin{eqnarray}\label{A6}
&& \Gamma_{tx}^{y}=\Gamma_{xt}^{y}= - \Gamma_{ty}^{x}= - \Gamma_{yt}^{x}=\Omega, \nonumber\\  %&\quad& \Gamma_{ty}^{x}=\Gamma_{yt}^{x}=-\Omega,
&& \Gamma_{tt}^{x}=-\Omega^{2}x, \quad \Gamma_{tt}^{y}=-\Omega^{2}y,
\end{eqnarray}
the nonzero component of affine connection $\Gamma_{\mu}$ is given by $\Gamma_{t}=-\frac{i}{2}\Omega\sigma^{12}$.
%\begin{eqnarray}\label{A7}
%\Gamma_{t}=-\frac{i}{2}\Omega\sigma^{12},\quad\Gamma_{x}=\Gamma_{y}=\Gamma_{z}=0.
%\end{eqnarray}
At this stage, in order to bring the $\gamma$-matrices written in general coordinate in a more appropriate form, using $\gamma^{\mu}=e^{\mu}_{~a}\gamma^{a}$ with the nonvanishing elements of $e^{\mu}_{~a}$ given in \eqref{A5}, the $\gamma$-matrices in general coordinate are given by
\begin{eqnarray}\label{A8}
\begin{array}{rclcrcl}
\gamma^{t}&=&\gamma^{0},&\quad&\gamma^{x}&=& \Omega y \gamma^{0}+\gamma^{1},\\
\gamma^{y}&=&- \Omega x \gamma^{0}+\gamma^{2},&\quad& \gamma^{z}&=&\gamma^{3}.
\end{array}
\end{eqnarray}
Combining all results for nonvanishing elements of $\Gamma_{\mu}$ as well as $\gamma^{\mu}$ from \eqref{A8} and plug them into \eqref{A3} the Lagrangian density of NJL model in a rotating frame at finite density is, thus, given by
\begin{equation}\label{A9}
\mathcal{L} =  \bar \psi \left(\Pi + \mu \gamma^{0}  \right) \psi + \frac{G}{2} \left[ \left( \bar \psi \psi  \right)^{2} + \left( \bar \psi  i\gamma^{5}  \bm{\tau} \psi  \right)^{2}  \right],
\end{equation}
where the modified differential operator $\Pi$ is defined as
\begin{equation}\label{A10}
\Pi \equiv
i\gamma^{0}\left(\partial_{t}-i\Omega \hat{J}_{z}\right)+i\gamma^{1} \partial_{x} + i\gamma^{2} \partial_{y} + i\gamma^{3}\partial_{z},
\end{equation}
with $\hat{J}_{z}\equiv\hat{L}_{z}+\Sigma_{z}/2$, the total azimuthal angular momentum, $\hat{L}_{z}\equiv-i\left(x\partial_{y}-y\partial_{x}\right)$, the orbital angular momentum, and the spin along the $z$-direction $\Sigma_{z} = i \gamma^{1}\gamma^{2}$.
\par
Concerning the thermodynamic potential of corresponding Lagrangian \eqref{A10}, following the standard procedure to bosonize the model by introducing mesonic auxiliary fields $(\sigma , \bm{\pi})$ (see Appendix \ref{appendixA} for more details), the semibosonized Lagrangian density takes the form
\begin{eqnarray}\label{A11}
\mathcal{L}_{SB} = && - \frac{\sigma^{2} + \bm{\pi}^{2}}{2G} \nonumber \\
&& + \bar \psi \left ( \Pi - \sigma - i \gamma^{5} \bm{\tau} \cdot \bm{\pi} + \mu \gamma^{0} \right ) \psi,
\end{eqnarray}
where the spatial modulation of scalar $\sigma$ and pseudoscalar $\bm{\pi}$ fields, according to dual chiral density wave, in the mean field approximation are defined by
\begin{eqnarray}\label{A12}
\sigma &\equiv& -G \langle \bar \psi \psi \rangle = m \cos (q z), \quad \pi_{1}=\pi_{2}=0, \nonumber \\ 
\pi^{0} &\equiv& -G \langle \bar \psi i \gamma^{5} \tau_{3} \psi \rangle = m \sin (q z).
\end{eqnarray}
Here, the constituent mass $m=- G \Delta$, the chiral density amplitude $\Delta$, and the wave vector $q$ whose magnitude controls the degree of spatial inhomogeneity of condensate. It should be noted that the condensate pair of particles in the DCDW form a standing wave with nonvanishing momentum. Moreover, both $\sigma$ and $\pi^{0}$ fields lie on the circle of constant radius $m$. For the special case of a vanishing wave vector, the DCDW reduces to the conventional homogeneous scalar chiral condensate. Both parameters are generated dynamically, thus, they have to be determined by minimizing the corresponding thermodynamic potential.
\par
At this stage, in order to obtain the thermodynamic potential, after performing the integral over fermionic degrees of freedom as well as following the steps in Appendix \ref{appendixA}, the thermodynamic potential $\Omega_{\text{eff}} = \Omega^{(0)}_{\text{eff}} + \Omega^{(1)}_{\text{eff}}$ of the two-flavor NJL model at finite temperature and density in the corotating frame is given by
\begin{widetext}
\begin{eqnarray}\label{A13}
\Omega_{\text{eff}} &=& \frac{1}{V} \int d^{3} x \left [ \frac{m^{2}}{2 G} + \frac{N_{f} N_{c}}{8 \pi^{5/2}} \sum_{s = \pm} \int_{\Lambda^{-2}}^{\infty} \frac{d \tau}{ \tau^{5/2}} \int_{0}^{\infty} dk_{z}  \exp \left[ - \left( \sqrt{k_{z}^{2}+m^{2}} +  \frac{s \, q}{2} \right)^{2}  \tau \right] - \frac{N_{f} N_{c} T}{4 \pi^{2}} \sum_{s = \pm} \sum_{\ell = -\infty}^{\infty} \int_{0}^{\infty} dk_{z} \times     \right. \nonumber \\
&& \left. \int_{0}^{\infty}  dk_{\perp}  k_{\perp}  \left( J^{2}_{\ell} (k_{\perp} r) + J^{2}_{\ell + 1} (k_{\perp} r) \right)  \left \{ \ln \left( 1 + e^{- \beta (\varepsilon_{s} + \Omega (\ell + 1/2) + \mu )}  \right) + \ln \left( 1 + e^{- \beta ( \varepsilon_{s} - \Omega (\ell + 1/2) - \mu )}  \right) \right \} \right ],
\end{eqnarray}
\end{widetext}
where $V$ and $J_{\ell} (x) $ denote the three-dimensional volume and the Bessel function of the first kind, respectively. Moreover, the dispersion relation defined as
\begin{eqnarray}\label{A14}
\varepsilon^{2}_{s} = k^{2}_{\perp} + \left( \sqrt{k_{z}^{2} + m^{2}} + s \frac{q}{2} \right)^{2}.
\end{eqnarray}
Here, the subscript $s = \pm 1$ refers to the spin direction. As it turns out, the thermodynamic potential $\Omega_{\text{eff}}$ in \eqref{A13} is invariant under the transformation $\Omega \rightarrow - \Omega$; thus, the direction of rotation e.g. clockwise or counterclockwise gives the same results. Let us note that the angular velocity in \eqref{A13} appears as a Lagrange multiplier of total angular momentum whose effect imitates the role of the chemical potential. In what follows, we scrutinize the similarities, differences, and the interplay of chemical potential and angular velocity on the formation of an inhomogeneous phase of quark matter.
\par 
At this stage, let us emphasize that for a rotating fermionic system, in particular, there is a freedom to define the vacuum state and correspondingly the particle and antiparticle modes. As it turns out, in the case of fermionic fields, this freedom is related to the positive norm of modes \cite{Ambrus:2014uqa}. Therefore, choosing an appropriate vacuum state is crucial to studying the thermodynamics of a system. In particular, two common choices are either the Minkowski vacuum, with $\varepsilon_{\pm} > 0$, defined in \cite{Vilenkin:1980zv} or the corotating vacuum state, with $\varepsilon_{\pm} - \Omega (\ell + 1/2) > 0$, introduced in \cite{Iyer:1982ah}. Thus, to fix the vacuum state, we choose the former definition to perform our numerical analysis. Moreover, noting in particular that, in a rotating system, the constituent mass is inhomogeneous and depends on the radial distance from the axis of rotation, we obtain $\Omega_{\text{eff}}$ in \eqref{A13} by employing the so-called local density approximation $\partial_{r} m \ll m^{2}$ \cite{Sadooghi:2021upd,Ebihara:2016fwa}. In Sec. \ref{NumRes}, we use $\Omega_{\text{eff}}$ in \eqref{A13} to determine the dependence of constituent mass $m$ and the wave vector $q$ on chemical potential and angular velocity at zero and finite temperature. Then, we present the phase diagram of the rotating two-flavor quark matter in the NJL model at finite $T$, $\mu$, and $\Omega$.
%%%%%%%%%%%%%
\subsection{Solution of Dirac equation} \label{sec2B}
%%%%%%%%%%%%%
In this section, we find the solution of the Dirac equation in a rotating frame with an inhomogeneous chiral condensate. To include the effect of DCDW, let us start with the semibosonized Lagrangian density of the fermionic field in the mean-field approximation
%%%%
\begin{figure*}[hbt]
\includegraphics[width=8cm,height=6.5cm]{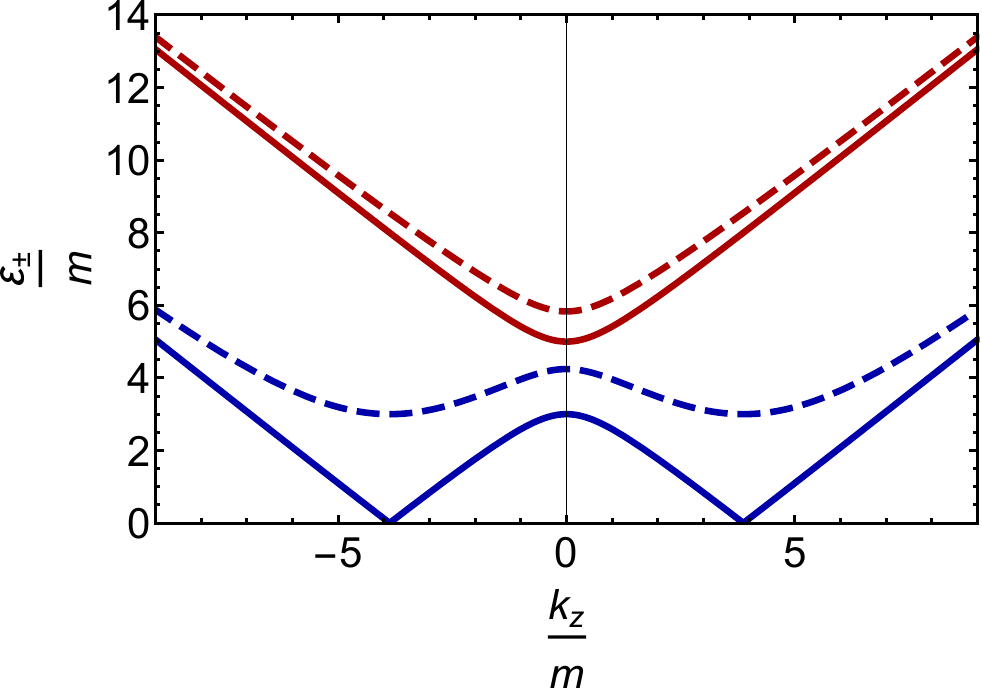} \hspace{0.2cm}
%\vskip 0.1 cm
\includegraphics[width=8cm,height=6.5cm]{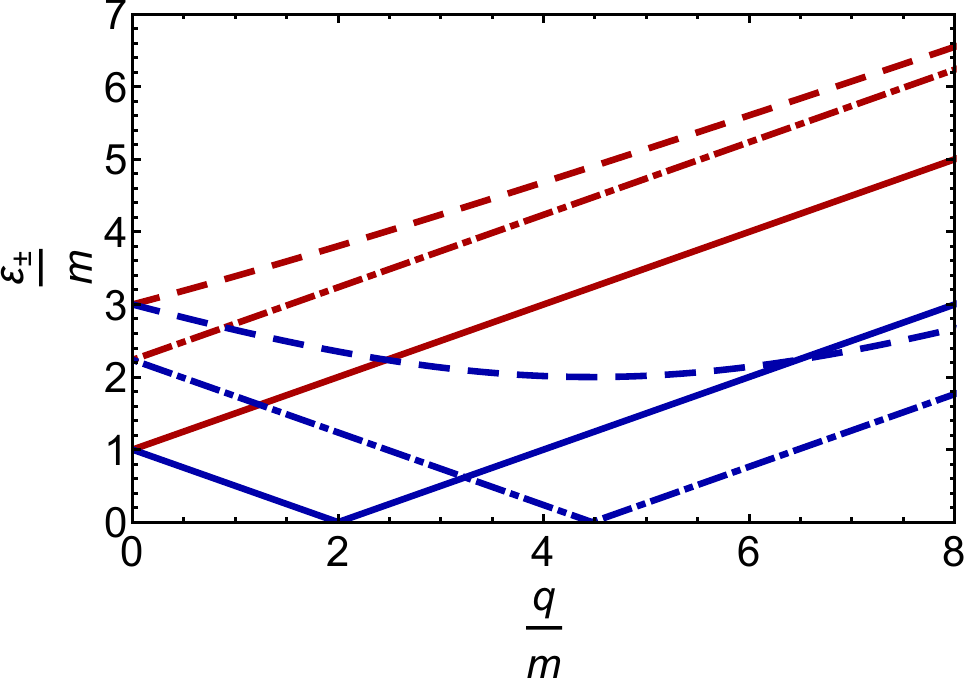}
\caption{(color online). In the left panel, the $k_{z}$ dependence of two branches of dispersion relation $\varepsilon_{+}$ (red lines) and $\varepsilon_{-}$ (blue lines) plotted for $q = 8 m$. Here, the solid (dashed) lines denote the $k_{\perp} =0$ ($k_{\perp} = 3 m$). In the right panel, the wave vector $q$ dependence of two branches of the dispersion relation is plotted. Here, the color code for $\varepsilon_{\pm}$ is the same as the left panel. Furthermore, the solid, dot-dashed as well as the dashed lines correspond to $(k_{z}=k_{\perp} =0)$, $(k_{z}=2m, k_{\perp}=0)$ and $(k_{z} = k_{\perp} = 2m)$, respectively. }\label{energ}
\end{figure*}
%%%%
\begin{eqnarray} \label{A15}
\mathcal{L}_{SB} = \bar \psi \left( \Pi - \sigma - i \gamma^{5} \bm{\tau} \cdot \bm{\pi} \right) \psi,
\end{eqnarray}
where $\Pi$ defined in \eqref{A10}. To elaborate the effect of rotation on the formation of inhomogeneous phase as well as the neutral pion $\pi^{0}$ condensate as a by-product, we assume the $\sigma$ and $\pi^{0}$ fields take the following form
\begin{equation}\label{A16}
\langle \bar \psi \psi \rangle =  \Delta \cos \vartheta, \quad  \langle \bar \psi i \gamma^{5} \tau_{3} \psi \rangle =  \Delta \sin  \vartheta,
\end{equation}
where the angle $\vartheta$ depends on the spatial coordinate. Plugging, at this stage, the scalar and pseudoscalar condensate in \eqref{A16} into the Lagrangian density \eqref{A15}, as well as performing the chiral transformation\footnote{In the absence of electromagnetic field this is justified. In the presence of an electromagnetic field, the transformation is to be done using the path integral.} $\psi \rightarrow \psi_{w} \equiv e^{-i \tau_{3} \gamma^{5} \vartheta /2} \psi$, the Hamiltonian of system reads
\begin{eqnarray}\label{A17}
	\mathcal{\hat H}_{w} &=& - i \gamma^{0} \left ( \gamma^{1} \partial_{x} + \gamma^{2} \partial_{y} + \gamma^{3}\partial_{z} \right) + m \gamma^{0} + \Omega \, \hat{J}_{z}  \nonumber \\
	&& + \frac{i}{2} \gamma^{5} \gamma^{0} \tau_{3} (\mathcal{D} \vartheta)  - \frac{i}{2} \gamma^{5} \tau_{3} \Omega (\hat{L}_{z} \vartheta),
\end{eqnarray}
where $\Pi = \mathcal{D} + \gamma^{0} \Omega \, \hat{J}_{z}$. At this stage, let us emphasize that in a rotating frame, the inhomogeneous chiral condensate in the transverse plane relative to $\Omega$ arises naturally \cite{Jiang:2016wvv,Sadooghi:2021upd}. Moreover, assuming $\vartheta = \textbf{b} \cdot \textbf{x}$ it was shown in \cite{Tatsumi:2003bq} the spin of each flavor is polarized along the direction of wave vector with a different sign. Bearing in mind that in a rotating system, the spin is aligned along the $\Omega$ axis, we thus consider the wave vector as $\textbf{b} = (0,0,q)$.
\par
In the rest of this section, we find the solution of one flavor Dirac equation whose explicit form is given by
\begin{eqnarray} \label{A18}
\left( \Pi - m - \frac{q}{2} \gamma^{5} \gamma^{3} \right) \psi^{(s)}_{w} =0,
\end{eqnarray}
with the superscript $s = \pm$. Noting that the matrix structure of the last term in the Dirac equation is similar to the spin-angular velocity coupling in operator $\Pi$ defined in \eqref{A10}, therefore, this coupling has a renormalization effect on the wave vector. Investigating the effect of this coupling on the phase diagram of this model is the main purpose of this paper.
\par
The solution of the Dirac equation in \eqref{A18} is determined by constructing the simultaneous eigenfunction of commuting operators, azimuthal angular momentum $[ \mathcal{\hat H}_{w},\hat{J}_{z}]=0$ whose eigenvalues are given by $\hat{J}_{z} \psi^{(s)}_{w} = (\ell + 1/2) \psi^{(s)}_{w}$, the transverse momentum $[ \mathcal{\hat H}_{w} , \hat K_{T}^{2}] =0 $ with $\hat K_{\perp}^{2} \psi^{(s)}_{w} = k_{\perp}^{2} \psi^{(s)}_{w}$ as well as third component of momentum operator $\hat{K}_{z} \psi^{(s)}_{w} = k_{z} \psi^{(s)}_{w}$. Here, after plugging the wave vector $\textbf{b} = (0,0,q)$ into the \eqref{A17}, the corotating Hamiltonian reduces to
\begin{eqnarray}\label{A19}
\mathcal{\hat H}_{w} = && -i \gamma^{0} \left( \gamma^{1} \partial_{x} + \gamma^{2} \partial_{y} + \gamma^{3}\partial_{z} \right) + \Omega \, \hat{J}_{z}  \nonumber \\
&&  - \frac{q}{2} \Sigma_{z} + m \gamma^{0}.
\end{eqnarray}
At this stage, using the Weyl representation of the $\gamma$-matrices as well as employing the chiral decomposition of the wave function
\begin{eqnarray}\label{A20}
\psi^{(s)}_{w} = \mathcal{N} e^{- i E t + i k_{z} z} \left( \begin{array}{cc}
\psi^{(s)}_{L} \\
\psi^{(s)}_{R}
\end{array}
\right)
\end{eqnarray}
the differential equation for the left- and right-handed spinors read
\begin{eqnarray}\label{A21}
&&\left( \mathcal{D}_{R} + \frac{q}{2} \sigma_{z} \right) \psi^{(s)}_{R} - m \psi^{(s)}_{L} =0, \nonumber \\
&& \left( \mathcal{D}_{L} + \frac{q}{2} \sigma_{z} \right) \psi^{(s)}_{L} - m \psi^{(s)}_{R} =0,
\end{eqnarray}
where 
\begin{eqnarray}\label{A22}
\mathcal{D}_{L/R} = && E + \Omega (\hat L_{z} + \frac{1}{2} \sigma_{z}) \nonumber \\ &&  \mp i \left( \sigma_{x} \partial_{x} +  \sigma_{y} \partial_{y} + i \sigma_{z} k_{z} \right ).
\end{eqnarray}
Let us note that, $\mathcal{N}$ in \eqref{A20} is the normalization factor. Moreover, the field $\psi^{(s)}_{w}$  is the eigenvector of $\hat{J}_{z}$, therefore its angular dependence is given according to
 \begin{eqnarray} \label{A23}
 \psi^{(s)}_{a} = \left(
\begin{array}{cc}
e^{i \ell \phi} \psi_{a,+} \\
e^{i (\ell + 1) \phi} \psi_{a,-}
\end{array}
\right),  \qquad a=L,R.
\end{eqnarray}
Plugging \eqref{A23} into \eqref{A21} we arrive at 
\begin{widetext}
\begin{eqnarray}\label{A24}
\left( (E + \Omega j + \frac{q}{2})^{2} - k_{z}^{2} - m^{2} + \partial_{r}^{2} + \frac{1}{r} \partial_{r} - \frac{\ell^{2}}{r^{2}} \right) \psi_{L,+} - i q \left( \partial_{r} + \frac{\ell + 1}{r} \right) \psi_{L,-} = 0, \nonumber \\
\left( (E + \Omega j - \frac{q}{2})^{2} - k_{z}^{2} - m^{2} + \partial_{r}^{2} + \frac{1}{r} \partial_{r} - \frac{( \ell+1)^{2}}{r^{2}} \right) \psi_{L,-} + i q \left( \partial_{r} - \frac{\ell}{r} \right) \psi_{L, +} = 0.
\end{eqnarray}
\end{widetext}
At this stage, combining the above equations to arrive at an equation for $\psi_{L,\pm}$ we obtain $E_{s} +  \Omega j = \varepsilon_{s}$ with $\varepsilon_{s}$ given by \eqref{A14}. As it turns out, one of the nontrivial consequences of the spatial dependence of chiral condensate, whose modulation is described according to DCDW, is the change in the energy dispersion and consequently the Fermi surface. In particular, the dual chiral density wave induces an effective interaction of axial current $\bar \psi \gamma^{5} \gamma^{i} \psi$, whose role as the spin density resolves the spin degeneracy of spin degrees of freedom, with the gradient of modulation $\nabla_{i} \vartheta$. In the left panel of Fig. \ref{energ}, the dispersion relation $\varepsilon_{+}$ (red line) and $\varepsilon_{-}$ (blue line) demonstrated for $k_{\perp} = 0$ (solid line), $k_{\perp} = 3m$ (dashed line) and $q= 8 m$. As it is shown, comparing two branches of energy $\varepsilon_{+}$ and $\varepsilon_{-}$, we observe the latter has a lower-energy level due to the nonvanishing wave vector. To scrutinize this point, in the right panel of this figure, we plot the wave-vector dependence of $\varepsilon_{\pm}$. As demonstrated, $\varepsilon_{+}$ increases by increasing the wave vector whereas the second branch of dispersion relation $\varepsilon_{-}$ has an opposite tendency and decreases with increasing the $q$. Therefore, particles fill this branch of energy as it lowers the energy of the system.
\par
Regarding the fermionic wave function, in order to solve the differential equations in \eqref{A24}, we note that for a particular case of $q=0$ the regular solutions at $r \rightarrow 0$ are given by
\begin{equation}\label{A25}
\psi_{L,+} = \mathcal{A}^{+} \, J_{\ell} (k_{\perp} r), \quad \psi_{L,-} = \mathcal{A}^{-} \, J_{\ell + 1} (k_{\perp} r),
\end{equation}
where $\mathcal{A}^{\pm}$ are the normalization factors which will be determined subsequently. Plugging, \eqref{A25} into \eqref{A24} the corresponding solution of left-handed wave function reads
\begin{eqnarray*}\label{A26}
\psi_{L} =  \mathcal{B}  \left( 
\begin{array}{c}
i \left( \varepsilon_{s} - \frac{q}{2} - s \sqrt{m^{2} + k_{z}^{2}} \right) e^{i \ell \phi}  J_{\ell} (k_{\perp} r) \\
k_{\perp}  e^{i (\ell + 1) \phi} J_{\ell + 1} (k_{\perp} r)
\end{array}
\right),
\end{eqnarray*}
with $\mathcal{B} = \left( \varepsilon_{s} - \frac{q}{2} - s \sqrt{m^{2} + k_{z}^{2}} \right)^{-1/2}$. As concerns the right-handed solution of the Dirac equation, we use the second equation of \eqref{A21}. After some lengthy but straightforward manipulation, this equation is written as
\begin{eqnarray}\label{A27}
\psi^{(s)}_{R} = - \frac{m}{k_{z} + s \sqrt{k_{z}^{2} + m^{2}}} \sigma_{z} \psi^{(s)}_{L}.
\end{eqnarray}
Finally, Noting that in the curved space-time, the wave function satisfies the orthogonality condition \cite{Iyer:1982ah}
\begin{eqnarray}\label{A28}
\langle \psi , \phi \rangle &\equiv& \int d^{3}x \sqrt{-g} \bar \psi \gamma^{t} \phi  \nonumber \\ 
&=&  2 \pi \delta_{\ell \ell{\prime}} \delta (k_{z} - k_{z}^{\prime}) \frac{\delta(k_{\perp} - k_{\perp})}{k_{\perp}} ,
\end{eqnarray}
where for the metric \eqref{A1}, the normalization factor of the wave function \eqref{A20} is given by
\begin{eqnarray}\label{A29}
\mathcal{N} = \frac{1}{2} \left( \frac{k_{z} + s \sqrt{m^{2} + k_{z}^{2}} }{ \varepsilon_{s}  s \sqrt{m^{2} + k_{z}^{2}}} \right)^{1/2}
\end{eqnarray}

%%%%%%%%%%%%%
\section{NUMERICAL RESULTS} \label{NumRes}
\setcounter{equation}{0}
%%%%%%%%%%%%%
In Sec. \ref{sec2A}, we determined thermodynamic potential $\Omega_{\text{eff}}$ of the NJL model developing DCDW in the mean-field approximation at finite $T$, $\mu$, and $\Omega$. We aim, in this section, to present the numerical results on the $\mu$, $T$, and $\Omega$ dependence of dynamically generated quantities, the constituent mass, and the wave vector. To do this, we divide this section into two parts. First, we focus on the zero-temperature limit. Therefore, after presenting the numerical results on the $\mu$ dependence of $m$ and $q$ at various $\Omega$, we shed light, in particular, on new aspects that were not discussed before. We report the possibility of having DCDW even at small $\mu$. Then, to capture all this information, we further provide the phase structure of DCDW in the $\mu - \Omega$ phase plane. Then, in the second part of this section, we study our model at finite temperature and present the $T - \mu$, $T - \Omega$ as well as $\mu - \Omega$ phase portraits of two-flavor rotating quark matter.
%%%%
\begin{figure*}[hbt]
\includegraphics[width=8cm,height=6.5cm]{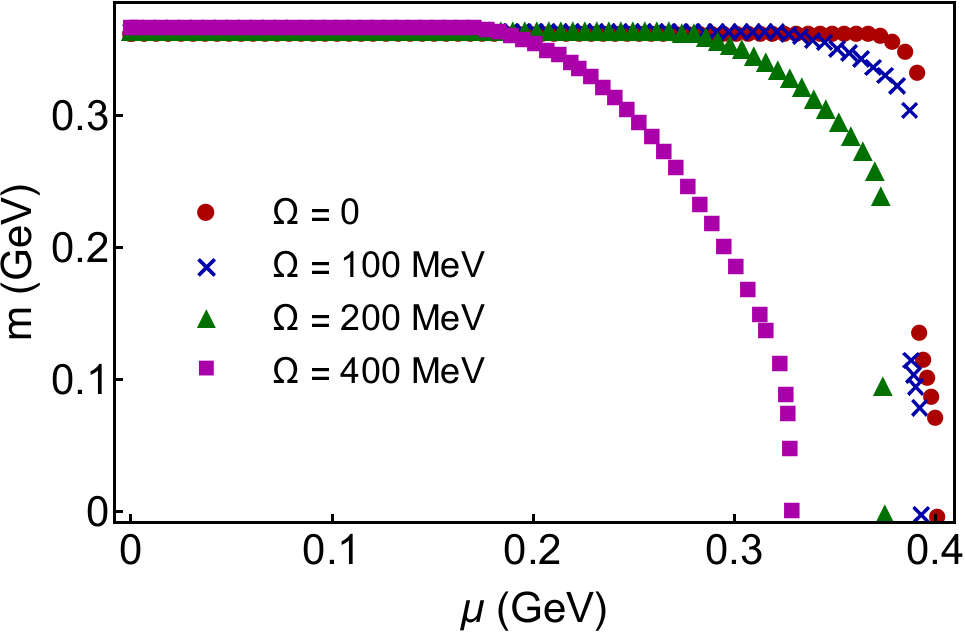} \hspace{0.2cm}
%\vskip 0.1 cm
\includegraphics[width=8cm,height=6.5cm]{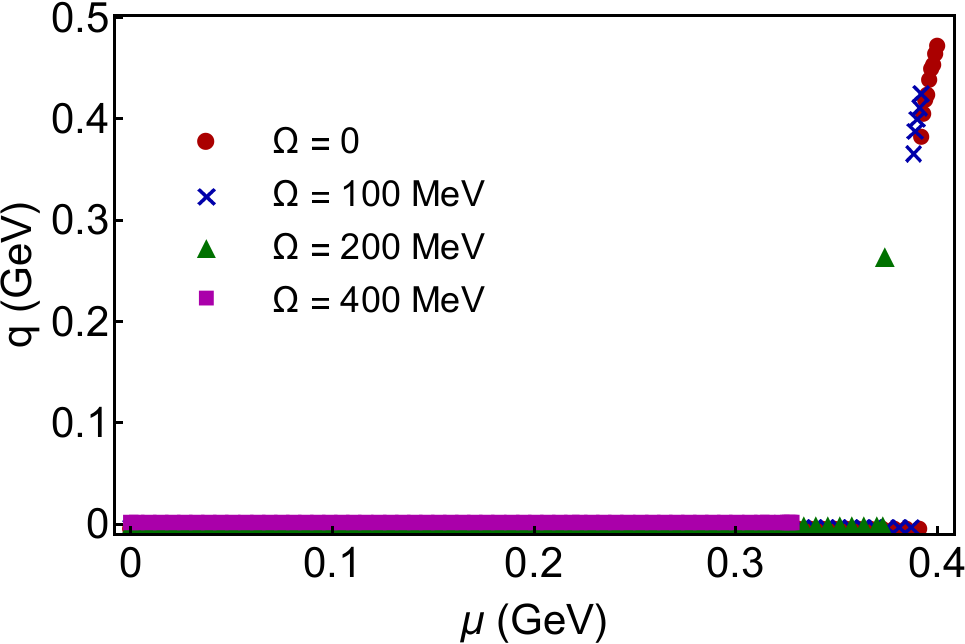}
\caption{(color online). The $\mu$ dependence of the constituent mass $m=-G\Delta$ (left panel) and the wave vector $q$ (right panel) for different values of angular velocity ($\Omega = 0, 100, 200, 400 $ \text{MeV}). }\label{fig1}
\end{figure*}
%%%%
\par
At this stage, let us start with the thermodynamic potential \eqref{A13}. The constituent mass $m$ and the wave vector $q$ are dynamically generated. To determine these dynamical quantities, the corresponding  $\Omega_{\text{eff}}$ is to be minimized. Thus, the corresponding gap equations are given by
\begin{eqnarray}\label{B1}
\frac{\partial \Omega_{\text{eff}} (\mu, \Omega; m, q)}{\partial m(r)} = 0,
\end{eqnarray}
and
\begin{eqnarray}\label{B2}
\frac{\partial \Omega_{\text{eff}}(\mu, \Omega; m, q)}{\partial q} = 0.
\end{eqnarray}
In order to solve these gap equations, the parameters of the model, the cutoff $\Lambda$ and the NJL coupling constant $G$, are to be fixed. We find the global minimum of $\Omega_{\text{eff}}$ for a two-flavor $N_{f}=2$ quark matter with the number of color $N_{c} = 3$ and 
\begin{eqnarray} \label{B3}
\Lambda = 0.86 ~ \mbox{GeV}, \quad G \Lambda^{2} = 11.
\end{eqnarray}
Apart from the parameters of the model, according to Tolman-Ehrenfest law, in curved space-time, the temperature $T ({\bf{x}})$ of a system in thermal equilibrium is a local quantity. However, the product $T ({\bf{x}}) \sqrt{g_{00}} = T_{0}$ is constant and independent of spatial coordinates \cite{Tolman:1930zza,Tolman:1930ona}. In other words, for a rotating system, specifically, the temperature of quark matter at each point $T({\bf{x}})$, is related to the temperature at the axis of rotation $T_{0}$ according to $T(r) \sqrt{1-r^{2}\Omega^{2}}= T_{0}$. Thus, to simplify further, we perform the numerical solution at a particular distance from the axis of rotation $r=0.1 ~\mbox{GeV}^{-1}$ where $T(r) \approx T_{0}$.
%%%%%%%%%%%%%
\subsection{Zero temperature} \label{sec3A}
%%%%%%%%%%%%%
Bearing in mind that, the main purpose of this section is to study inhomogeneous chiral condensate of rotating quark matter at $T=0$, we  take the $T \rightarrow 0$ limit by applying the identity 
\begin{eqnarray}\label{B5}
\lim\limits_{\beta \rightarrow \infty} \frac{1}{\beta} \log \left ( 1 + e^{- \beta x} \right ) = -x \theta (-x),
\end{eqnarray}
with $\theta (x)$ is the Heaviside $\theta$-function. Following the above recipe, the one-loop $\Omega_{\text{eff}} = \Omega^{(0)}_{\text{eff}} + \Omega^{(1)}_{\text{eff}}$ at $T=0$ reads
\begin{widetext}
\begin{eqnarray}\label{B6}
\Omega_{\text{eff}} &=& \frac{1}{V} \int d^{3} x \Big [ \frac{m^{2}}{2 G} + \frac{N_{f} N_{c}}{8 \pi^{5/2}} \sum_{s = \pm} \int_{\Lambda^{-2}}^{\infty} \frac{d \tau}{ \tau^{5/2}} \int_{0}^{\infty} dk_{z}  \exp \left[ - \left( \sqrt{k_{z}^{2}+m^{2}} +  \frac{s \, q}{2} \right)^{2}  \tau \right] - \frac{N_{f} N_{c} T}{4 \pi^{2}} \sum_{s = \pm} \sum_{\ell = -\infty}^{\infty} \int_{0}^{\infty} dk_{z}   \times \nonumber \\
&& \int_{0}^{\infty}  dk_{\perp} k_{\perp} \left( J^{2}_{\ell} (k_{\perp} r) + J^{2}_{\ell + 1} (k_{\perp} r) \right)  \left \{ - (\mu + \Omega (\ell + 1/2) + \varepsilon_{s}) \theta (-\mu -\Omega (\ell + 1/2) - \varepsilon_{s})  \right. \nonumber \\
&&  \left. + (\mu + \Omega (\ell + 1/2) - \varepsilon_{s}) \theta \left(\mu + \Omega (\ell + 1/2) - \varepsilon_{s} \right) \right\}  \Big] \equiv \frac{1}{V} \int d^{3} x   \frac{m^{2}}{2 G} + \Omega^{(1)}_{\text{DS}} + \Omega^{(1)}_{\text{FS}}.
\end{eqnarray}
\end{widetext} 
%%%%
\begin{figure*}[hbt]
\includegraphics[width=5.7cm,height=5cm]{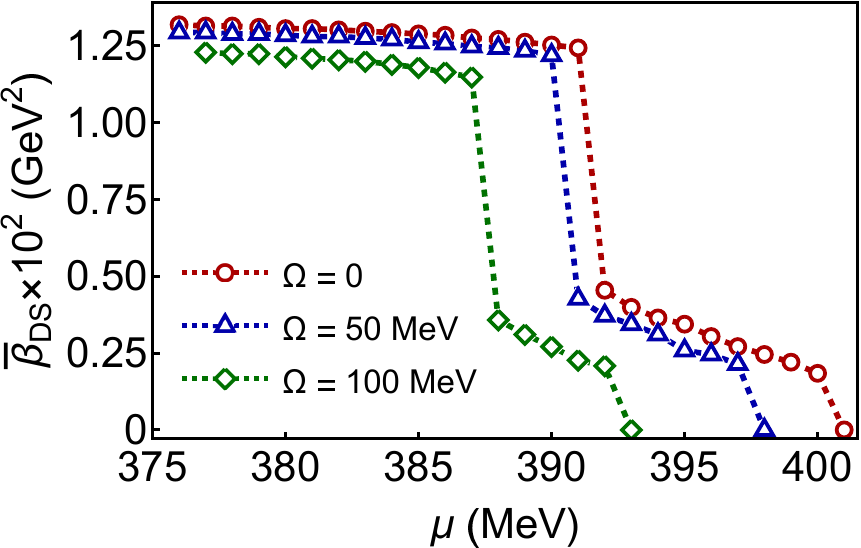}\hspace{0.05cm}
\includegraphics[width=5.7cm,height=5cm]{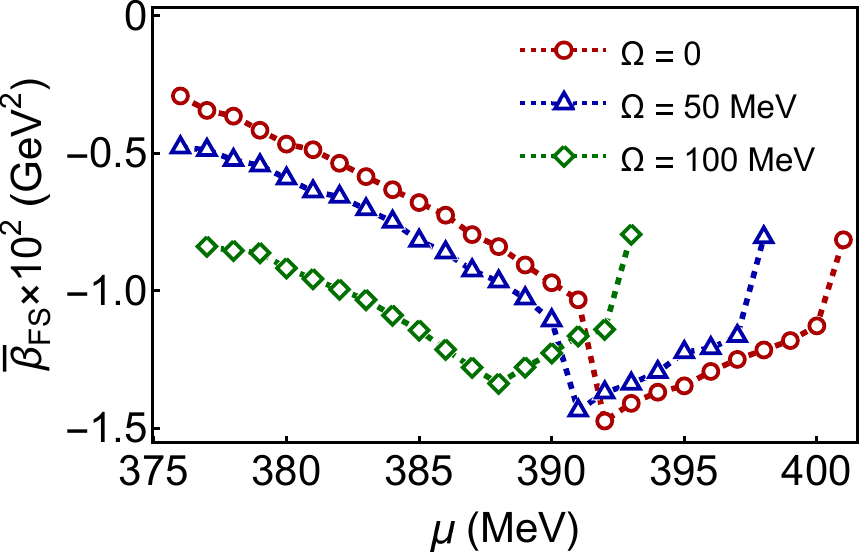} \hspace{0.05cm}
%\vskip 0.1 cm
%\includegraphics[width=8cm,height=6.5cm]{fig5c}\hspace{0.8cm}
\includegraphics[width=5.7cm,height=5cm]{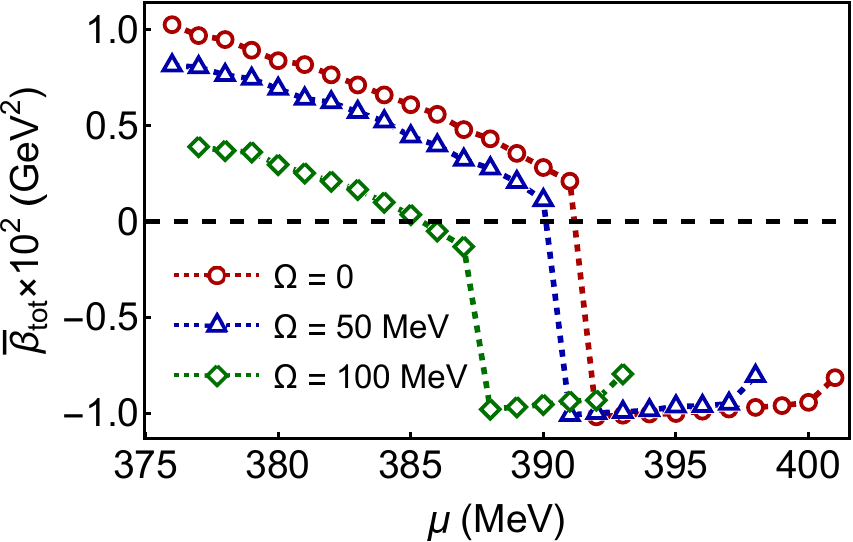}
\caption{(color online). The $\mu$ dependence of $\bar{\beta}_{DS}$ (left panel), $\bar{\beta}_{FS}$ (middle panel) and $\bar{\beta}_{tot}$ (right panel) at different angular velocity $\Omega = 0, 50, 100 \, \text{MeV}$. Despite $\bar{\beta}_{DS}$ has no explicit $\Omega$ dependence, an implicit dependence on angular velocity appears by the $\Omega$ dependence of constituent mass.}\label{BetaTot}
\end{figure*}
%%%%
Here, the one-loop correction part of thermodynamic potential $\Omega^{(1)}_{\text{eff}} = \Omega^{(1)}_{\text{DS}} + \Omega^{(1)}_{\text{FS}}$ contains two contributions from Dirac-sea (DS), or the vacuum part, and Fermi-sea (FS), respectively. It turns out that the rotation, in particular, changes the Fermi-sea contribution while keeping the vacuum part intact. 
\par
In Fig.\ref{fig1}, the dynamical mass (left panel) and wave vector (right panel) are plotted as the functions of $\mu$ for different values of angular velocity. As it turns out, at $\Omega = 0$, in particular, there are three distinct phases; the homogeneous $\chi SB$ phase, the inhomogeneous $\chi SB$ phase, and the $\chi RS$ phase. For the parameters used in this paper, the homogeneous phase exists for the range $0 \leq \mu \leq \mu_{1}$ where $\mu_{1} \approx 0.391 \text{GeV}$. At $\mu > \mu_{1}$, a first-order phase transition occurs from homogeneous phase \footnote{Let us recall that in a rotating system the constituent mass, still, has a mild radial dependence near the rotating axis in the local density approximation.} to the DCDW phase where nonvanishing $q$ arising in this state of quark matter. As it is shown in this figure, dynamical mass (wave vector) decreases (increases) with increasing $\mu$. Moreover, in this range of chemical potential, the strength of the wave vector is $q \approx \mathcal{O} ( (1-2) \mu)$. Increasing the $\mu$ further, at the critical value $\mu_{2} \approx 0.401 \text{GeV}$, the DCDW goes to the $\chi RS$ phase by a second-order transition. 
\par
%%%%
\begin{figure*}[hbt]
\includegraphics[width=8cm,height=6.5cm]{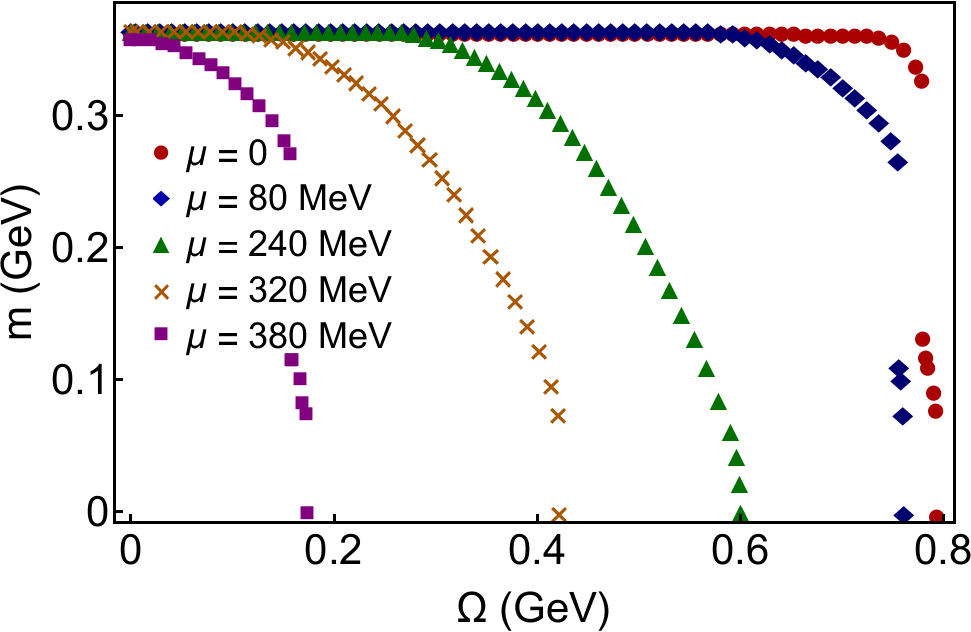} \hspace{0.2cm}
%\vskip 0.1 cm
\includegraphics[width=8cm,height=6.5cm]{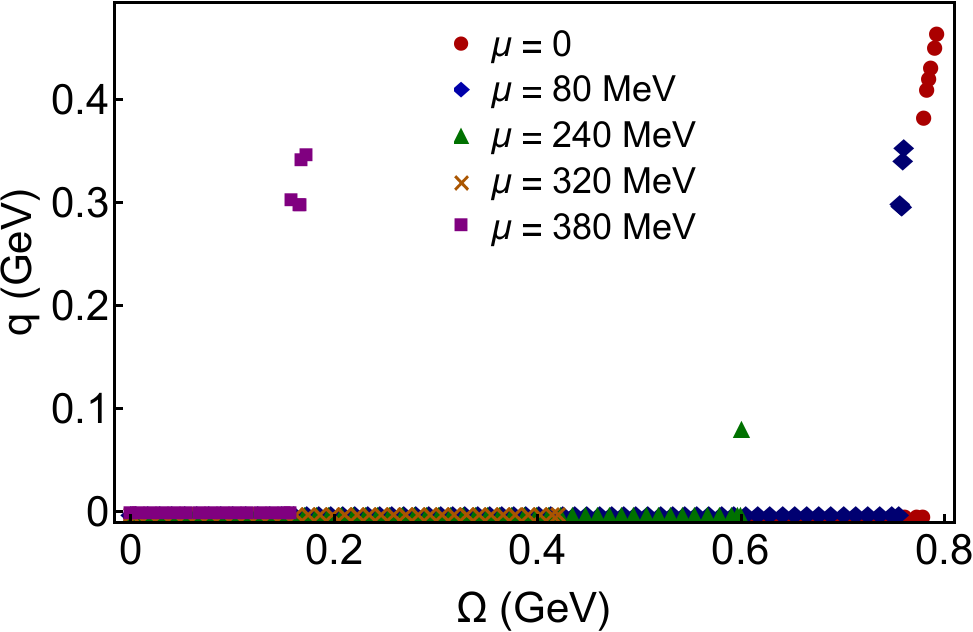}
\caption{(color online). The $\Omega$ dependence of the constituent mass $m=-G\Delta$ (left panel) and the wave vector $q$ (right panel) for various values of chemical potential ($\mu = 0, 80, 240, 320, 380$ \text{MeV}).  }\label{fig2}
\end{figure*}
%%%%
%%%%
\begin{figure*}[hbt]
\includegraphics[width=8cm,height=6.5cm]{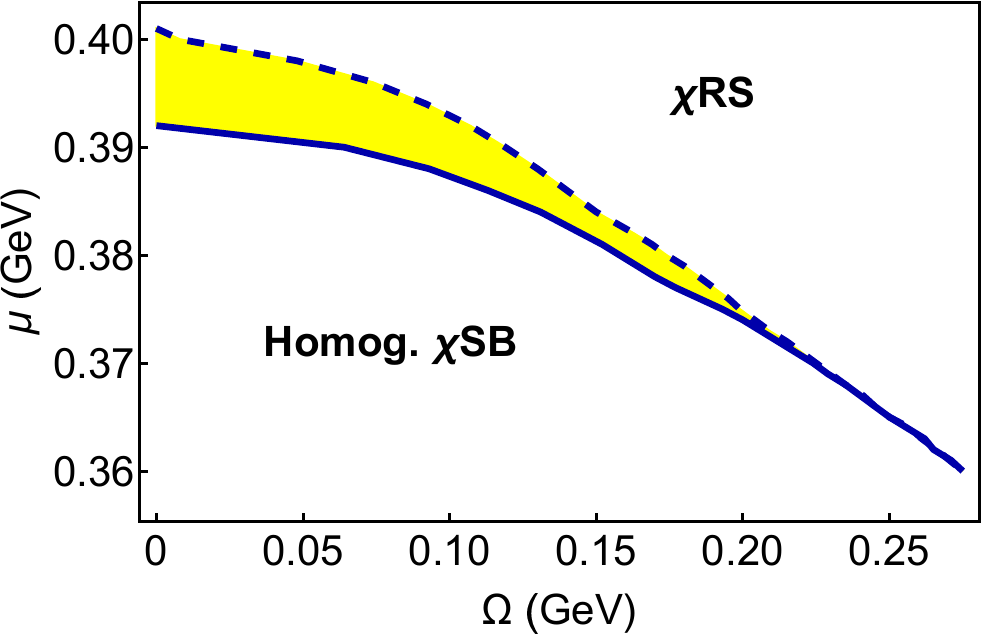}\hspace{0.8cm}
\includegraphics[width=8cm,height=6.5cm]{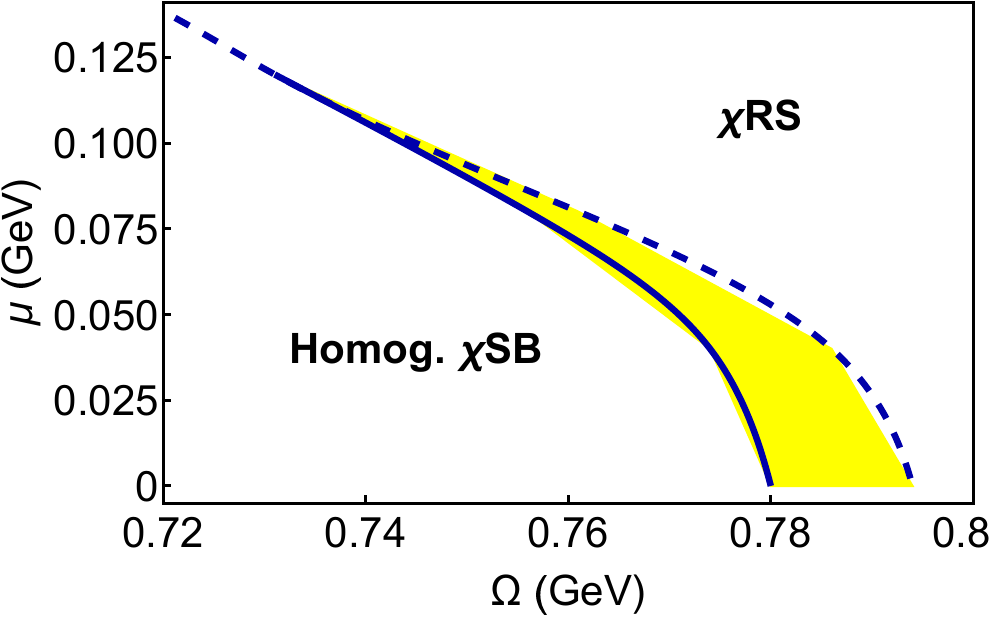}
\caption{(color online). The $\mu - \Omega$ phase diagram of a cold rotating two-flavor NJL model for two cases of small ratio of angular velocity to chemical potential $\frac{\Omega}{\mu} < 1$ (left panel) and large $\frac{\Omega}{\mu} > 1$ (right panel). The solid blue line denotes a first-order transition from the homogeneous phase to the inhomogeneous phase. This phase is shown by a yellow area in the phase plane. The dashed blue line indicates the second-order transition from inhomogeneous to chiral restored phase.}\label{fig3}
\end{figure*}
%%%%
As demonstrated in this figure, for nonvanishing $\Omega$, the interplay between the momentum modulation $q$ in the DCDW phase and angular velocity becomes relevant. For nonvanishing angular velocity, we observe three phases of quark matter. Comparing the behavior of $m$ as well as $q$ for different values of angular velocity, it turns out $\mu_{1}$ decreases to smaller values with increasing $\Omega$, leading to suppression of homogeneous phase. Moreover, by increasing the angular velocity the area of the inhomogeneous phase enclosed in range $ \mu_{1} < \mu < \mu_{2}$ starts to shrink. Thus, at some critical value $\Omega_{\text{crit}}$, we expect the DCDW phase to disappear. As it turns out, at angular velocity larger than $\Omega_{\text{crit}}$ and intermediate densities the transition from $\chi SB$ phase to $\chi RS$ is of second order.
\par
At this stage, using the solution of gap equations, we examine the effect of rotation on the Dirac and Fermi sea to assess the corresponding change in the threshold of forming the inhomogeneous chiral condensate. To do this, we expand the thermodynamic potential in \eqref{B6} around the homogeneous phase e.g., $\left ( \frac{\Omega_{\text{eff}} - \Omega_{\text{eff}} |_{q=0}}{\Lambda^{4}} \right) = \left ( \beta_{FS} + \beta_{DS} \right)\left( \frac{q}{\Lambda} \right )^{2} + \cdots$, with
\begin{eqnarray} \label{B6p5p5}
\beta_{DS} \Lambda^{2} &\equiv& \bar{\beta}_{DS} = \frac{\partial^{2} \Omega^{(1)}_{\text{DS}}}{\partial q^{2}}  \nonumber \\
&=&  \frac{N_{f} N_{c} \Lambda^{2}}{8 \pi^{2}} \left \{  (\frac{m}{\Lambda})^{2} \int_{(\frac{m}{\Lambda})^{2}}^{\infty} \frac{dt}{t} e^{-t} \right \},
\end{eqnarray}
and
%\begin{widetext}
\begin{eqnarray} \label{B6p5}
\hspace{-0.15 cm} \beta_{FS} \Lambda^{2} &\equiv& \bar{\beta}_{FS} = \frac{\partial^{2} \Omega^{(1)}_{\text{FS}}}{\partial q^{2}}  \nonumber \\
&= &  - \frac{N_{f} N_{c}}{4 \pi^{2}} \sum_{\ell = - \infty}^{\infty}  \int_{0}^{\sqrt{\left( \mu + \Omega (\ell + \frac{1}{2}) \right )^{2} - m^{2}}} dk_{\perp} k_{\perp} \nonumber \\
&& \left( J^{2}_{\ell} (k_{\perp} r) + J^{2}_{\ell + 1} (k_{\perp} r) \right) \left \{ 2 \frac{\left( \mu + \Omega (\ell + \frac{1}{2}) \right )^{2} - k_{\perp}^{2}}{\left( \mu + \Omega (\ell + \frac{1}{2}) \right )^{2}} \right. \nonumber \\
&&\left.  - \frac{k_{\perp}^{2} \sqrt{\left( \mu + \Omega (\ell + \frac{1}{2}) \right )^{2} - m^{2} - k_{\perp}^{2}}}{\left ( m^{2} + k_{\perp}^{2} \right) \left | \mu + \Omega (\ell + \frac{1}{2} \right| } \right \}.
\end{eqnarray}
%\end{widetext}
In Fig.\ref{BetaTot} we plot the chemical potential dependence of $\bar{\beta}_{DS}$ (left panel), $\bar{\beta}_{FS}$ (middle panel) and $\bar{\beta}_{tot} \equiv \bar{\beta}_{DS} + \bar{\beta}_{FS}$ (right panel). We observe the contribution of Dirac sea proportional to $\beta_{DS}$ tends to increase the free energy therefore it is against the formation of DCDW while the contribution of Fermi sea $\beta_{FS}$ supports the formation of DCDW by lowering the free energy. Once the quark matter starts to rotate, the angular velocity contribution appears in $\beta_{FS}$ whose effect prevails the Dirac sea at smaller values of chemical potential and consequently leads to the formation of DCDW. Let us note that, although $\bar{\beta}_{DS}$ in \eqref{B6p5p5} has no explicit dependence on the angular velocity, as shown in Fig.\ref{BetaTot}, obtains the corresponding dependence through the $\Omega$ dependence of constituent mass.
\par
As concerns the inhomogeneous phase in Fig.\ref{fig1}, it has been shown the existence of critical angular velocity $\Omega_{\text{crit}}$ above which the inhomogeneous phase is no longer favored. This observation indicates that DCDW appears in a finite area of the $\mu - \Omega$ plane. Thus, in order to scrutinize the interplay between angular velocity and DCDW further, we plotted in Fig.\ref{fig2} the $\Omega$ dependence of $m$ and $q$ for various $\mu$. As it turns out, at $\mu = 0$ an analogous effect to the conventional DCDW phase arises solely by rotation. Comparing this plot with the one in Fig.\ref{fig1}, it is obvious, that the inhomogeneous phase in this case arises at a larger value of angular velocity with approximate relation $\Omega \approx \mathcal{O} (2 \mu)$. This implies the wave vector strength is in the range $q \approx \mathcal{O} ((0.5-1) \Omega)$. Once the formation of an inhomogeneous phase at the rotating frame is established, increasing $\mu$ leads to suppression of $m$ as well as a smaller domain of DCDW wave. Thus, in analogy to the previous case, we anticipate at a critical chemical potential $\mu_{\text{crit}}$, the inhomogeneous phase no longer exists.
\par
%%%%
\begin{figure*}[hbt]
\includegraphics[width=5.7cm,height=5cm]{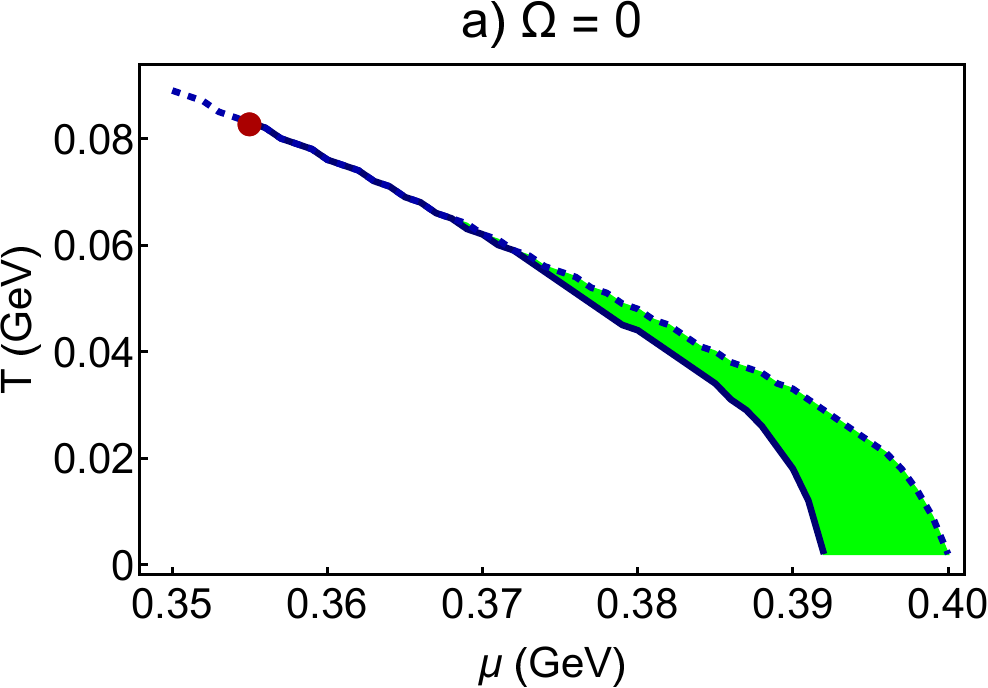}\hspace{0.05cm}
\includegraphics[width=5.7cm,height=5cm]{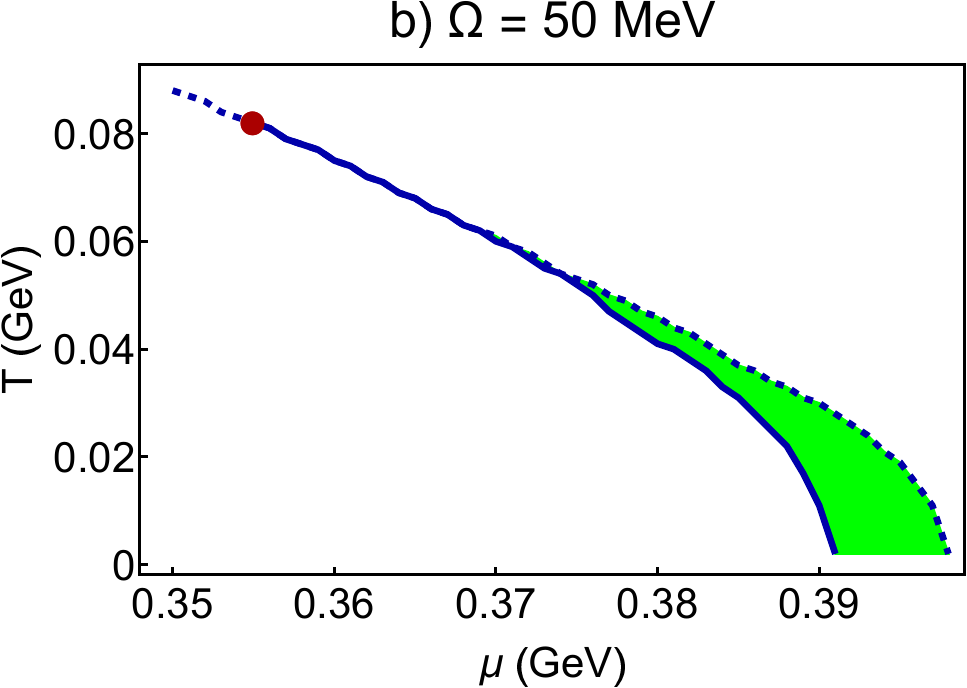} \hspace{0.05cm}
%\vskip 0.1 cm
%\includegraphics[width=8cm,height=6.5cm]{fig5c}\hspace{0.8cm}
\includegraphics[width=5.7cm,height=5cm]{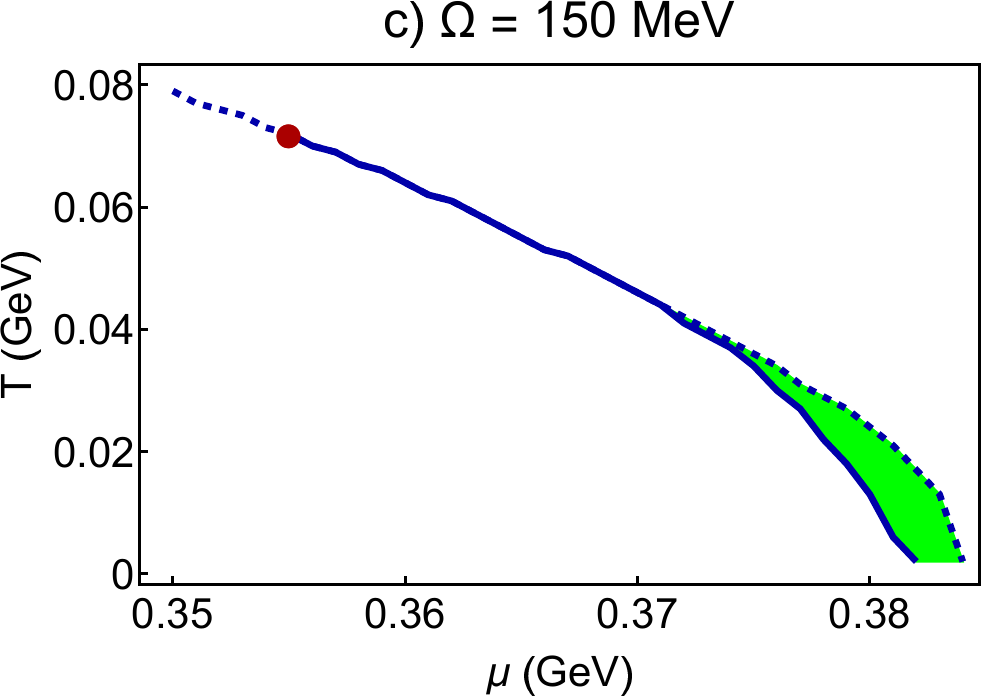}
\caption{(color online). The $T - \mu$ phase diagram of a hot-rotating quark matter for various values of $\Omega$. The blue solid line denotes the first-order transition from the homogeneous phase to the inhomogeneous one. The latter goes to the normal phase in a second-order transition shown by the blue dashed line. Here, the red bullet denotes the position of the critical point obtained by the Ginzburg-Landau analysis.}\label{phaseTMu}
\end{figure*}
%%%%
%%%%%%%%%
Combining the results obtained in Fig.\ref{fig1} and Fig.\ref{fig2}, we arrive at the phase diagram of DCDW for a cold-rotating quark matter. Thus, In Fig.\ref{fig3}, the $\mu - \Omega$ phase structure of the model is demonstrated. As we discussed earlier, three phases appear in this model, the homogeneous chiral broken phase, the inhomogeneous phase, and the normal ($\chi RS$) phase. As demonstrated in the left panel of this figure, for small $\frac{\Omega}{\mu} < 1$, the DCDW exists in a finite domain of $\mu$. In this plot, the transition from the homogeneous phase to the inhomogeneous one, the solid blue line, is of first order whereas the transition from DCDW to the normal state, the blue dashed line, is of second order. Moving to the higher angular velocities, along the horizontal axis, the DCDW is no longer favored over homogeneous or chiral restored phases. In this region of the $\mu - \Omega$ plane, the homogeneous phase goes to the restored phase by a second-order transition. However, as demonstrated in the right panel of Fig.\ref{fig3}, at the limit of ultrafast rotating quark matter $\frac{\Omega}{\mu} > 1$, the DCDW appears again. Similar to the former case, the transition from homogeneous to inhomogeneous is first order while from DCDW to restored one is of second order.

%%%%%%%%%%%%%
\subsection{The phase portrait of inhomogeneous chiral condensate of hot and rotating two-flavor NJL model} \label{sec3B}
\setcounter{equation}{0}
%%%%%%%%%%%%%
The aim of this paper is to study the effect of rotation on the inhomogeneous phase of quark matter. To this purpose, In Sec. \ref{sec2A}, we determined the thermodynamic potential of a two-flavor NJL model in a rotating frame. The former encodes the thermodynamic properties of the system under consideration and lets us study the phases of quark matter in a corotating frame. Then, In Sec.\ref{sec3A}, using the solution of gap equations for constituent mass and the wave vector, which also are subjected to be the global minimum of $\Omega_{\text{eff}}$, we studied the $\mu$- and $\Omega$-dependence of constituent mass and the wave vector at $T=0$.  In this section, we aim to generalize our previous results and explore the phase diagram of our model at finite temperature.
\par
At this stage, let us first consider $\Omega_{\text{eff}}$ in \eqref{A13}, at finite temperature and density in a rotating system. To explore the interplay between the chemical potential and the rotation of the system on the formation inhomogeneous phase, we first look at the solution of gap equations for various fixed $\Omega$. In Fig. \ref{phaseTMu}, the $T - \mu$ phase diagram of a hot-rotating two-flavor NJL model is plotted for $\Omega = 0, 50, 150$ \text{MeV}. In these plots, the solid blue line denotes the first-order transition, and the dashed blue line the second-order transition. Moreover, the red bullet in these figures denotes the location of the critical point in the phase plane. To find the former we perform a Ginzburg-Landau (GL) analysis and expand the thermodynamic potential $\Omega_{\text{eff}}$ in \eqref{A13} in terms of the order parameter $M = m e^{i q z}$ and its gradient as $\Omega_{\text{eff}} (T , \mu, \Omega; M) = \Omega_{\text{eff}} (T , \mu, \Omega; 0) + \alpha_{2} (T , \mu, \Omega) |M|^{2} + \alpha_{4,0} (T , \mu, \Omega) |M|^{4} + \alpha_{4,2} (T , \mu, \Omega) |\nabla M|^{2} + \cdots$ \cite{Nickel:2009ke,Nickel:2009wj}. For the particular case of DCDW, this approach is equivalent to the expansion of $\Omega_{\text{eff}}$ in terms of constituent mass $m$ and the wave vector $q$. According to the GL analysis the CP is given as the simultaneous solution of $\alpha_{2} = \alpha_{4,0} = 0$ \footnote{Let us note that, in the Ginzburg-Landau analysis of inhomogeneous phase of quark matter, due to the nonvanishing gradient of order parameter, it is possible to define the Lifshitz point (LP) by $\alpha_{2} = \alpha_{4,2} = 0$. It is argued in \cite{Nickel:2009ke,Nickel:2009wj}, that LP and CP coincide in the conventional NJL model with scalar and pseudoscalar channels. However, in this paper, for a system of rotating quark matter, we only examined the location of CP and skipped the relation between the LP and the CP in this system.}. 
%%%%
\begin{figure*}[hbt]
\includegraphics[width=5.7cm,height=5cm]{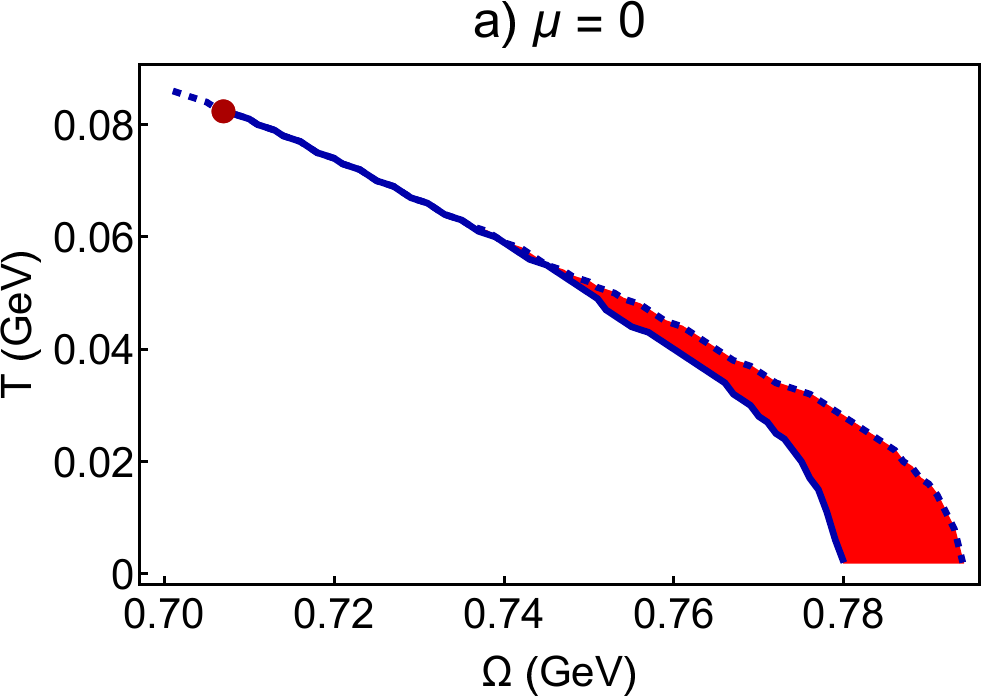}\hspace{0.05cm}
\includegraphics[width=5.7cm,height=5cm]{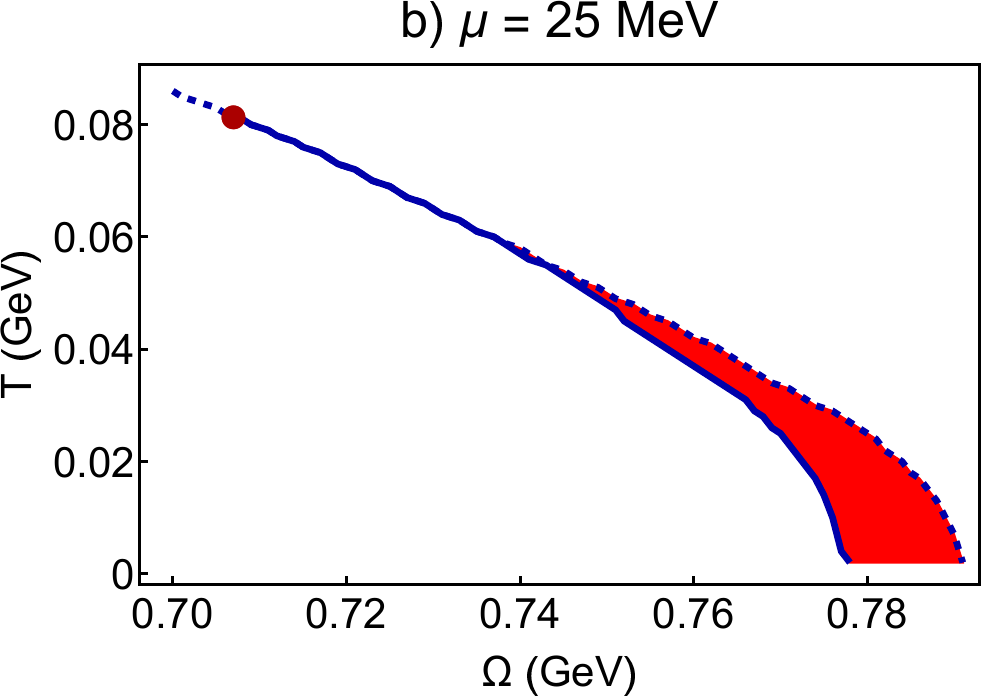} \hspace{0.05cm}
%\vskip 0.1 cm
%\includegraphics[width=8cm,height=6.5cm]{fig6c}\hspace{0.8cm}
\includegraphics[width=5.7cm,height=5cm]{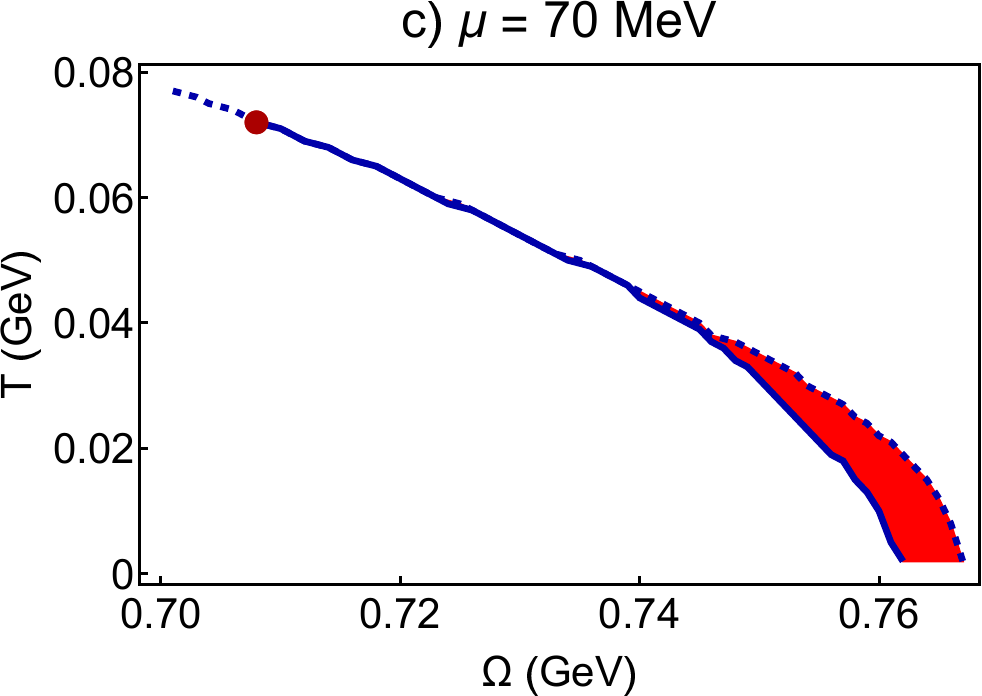}
\caption{(color online). The $T - \Omega$ phase diagram of a hot rotating two-flavor NJL model for different $\mu$. Here, the solid blue line is the first-order transition from homogeneous to DCDW whereas the dashed blue line denotes the second-order transition. Moreover, the red bullet, whose location in the phase portrait obtained by the Ginzburg-Landau analysis, denotes the critical point. }\label{phaseTOmega}
\end{figure*}
%%%%
\par
As it is shown, for a nonrotating system $\Omega = 0$ [see Fig.\ref{phaseTMu} (a)], three phases exist in the phase portrait. At small $\mu$ and $T$, the homogeneous phase is favored. By increasing the chemical potential $\mu > \mu_{1}$, after a first-order transition from $\chi SB$ to an inhomogeneous phase, the wave vector will have a nonvanishing value, leading to the formation of DCDW. At higher $\mu > \mu_{2}$, the DCDW goes to the chiral restored phase by a second-order transition. Once we increase the temperature, it turns out the thermal effects suppress the homogeneous chiral condensate which in turn, leads to entering the inhomogeneous phase at a smaller chemical potential $\mu_{1}$. On the other hand, similar to the former case, the amplitude of inhomogeneous chiral condensate decreases with temperature, thus, the inhomogeneous phase goes to the chiral restored phase at smaller $\mu_{2}$. This leads to the DCDW occurring in a smaller chemical potential interval as temperature increases.
\par
As concerned the effect of rotation, in Figs. \ref{phaseTMu}(b)$-$\ref{phaseTMu}(c), we plotted the $T - \mu$ phase portrait of rotating quark matter for $\Omega = 50, 150$ \text{MeV}, respectively. Comparing the nonrotating phase diagram with the rotating system, we observe the left and right sides of the inhomogeneous phase boundaries, corresponding to the first- and second-order transitions, moved to the region of the $T - \mu$ plane with smaller chemical potential. This originates from the suppression of chiral condensate by the rotation. On the other hand, the relevant chemical potential interval $\mu_{1} < \mu < \mu_{2}$, for the DCDW becomes smaller as the $\Omega$ increases. Moreover, as demonstrated in this plot, the length of the first-order phase boundary, in particular, decreases with increasing angular velocity. Hence, it is expected that at some critical $\Omega$, the phase diagram of our model exhibits no trace of DCDW, which in turn, the homogeneous phase goes to the $\chi RS$ phase by a second-order transition. Concerning the effect of $\Omega$ on the location of CP, in Table \ref{tab1} we presented the critical point $(\mu_{\text{crit}} , T_{\text{crit}} )$ obtained by the GL analysis. 
%%%%%%%
\begin{table}[hbt]
\caption{The location of critical points $(\mu_{\text{crit}} , T_{\text{crit}} )$ in the $T-\mu$ phase portrait of Fig.\ref{phaseTMu} for various values of angular velocity.}\label{tab1}
\vspace{1ex}
\begin{tabular}{ccc}
\hline \hline
$\Omega$ (MeV) $\qquad \qquad$ & $\mu_{\text{crit}}$(MeV) $\qquad \qquad$ & $T_{\text{crit}}$(MeV)\\
\hline
$0$ $\qquad \qquad$ &  $355$ $\qquad \qquad$ & $83$ \\
$50$ $\qquad \qquad$ &  $355$ $\qquad \qquad$ & $82$ \\
$150$ $\qquad \qquad$ &  $355$ $\qquad \qquad$ & $71$ \\
\hline \hline
\end{tabular}
\end{table}
%%%%%%%
It turns out, whereas, increasing the angular velocity keeps the position of the critical chemical potential $\mu_{\text{crit}}$ constant, it decreases the critical temperature $T_{\text{crit}}$ to smaller values.
\par
As it turns out, at regime $\frac{\Omega}{\mu}< 1$, the angular velocity changes the phase structure of DCDW drastically. At the higher densities which are comparable with the angular velocity, $\mu \approx \Omega$, there is no sign of DCDW where the homogeneous phase goes to the normal phase in a second-order transition. Thus, at this stage, to fill the missing gap of the phases of a hot-rotating quark matter at finite density, we focus on an ultrafast rotating system with $\frac{\Omega}{\mu} > 1$. \footnote{ Let us emphasize that to find the phases of matter in this limit, we do not employ any approximation. We work with the corresponding $\Omega_{\text{eff}}$ given by \eqref{A13}.} As discussed in Sec. \ref{sec3A}, in this domain of parameters, the system develops an inhomogeneous chiral condensate. Thus, it is our goal to scrutinize the fate of this phase at finite temperature. In Fig.\ref{phaseTOmega}, the phase diagram of DCDW in the $T - \Omega$ plane is demonstrated. As it is shown, at $\mu = 0$, the inhomogeneous chiral condensate starts in range $ \Omega_{1} < \Omega < \Omega_{2}$ with $\Omega_{1} = 0.78 \, \text{GeV}$ and $\Omega_{2} = 0.794 \, \text{GeV}$, respectively. By increasing the temperature, we observe a similar trend as the $T - \mu$ phase diagram, the phase boundaries move to regions of lower $\Omega$ and the difference $\Delta \Omega = \Omega_{2} - \Omega_{1}$ becomes smaller. Moving to the higher densities, we observe the phase boundaries move to low regions of angular velocity. This is similar to Fig.\ref{phaseTMu} where the increasing the $\Omega$ is accompanied by the transition to DCDW at a smaller chemical potential. Moreover, as demonstrated in Fig.\ref{phaseTOmega}, increasing the $\mu$ is accompanied by the decrease in the first-order phase boundary which leads to a smaller area of inhomogeneous phase. In order to observe the role of chemical potential on the position of CP, in Table \ref{tab2}, we listed the location of critical angular velocity $\Omega_{\text{crit}}$ and critical temperature $T_{\text{crit}}$ for various $\mu$. These critical points are denoted by a red bullet in Fig.\ref{phaseTOmega}. We observe, by increasing the chemical potential $T_{\text{crit}}$ decreases but $\Omega_{\text{crit}}$ is almost constant.
%%%%
\begin{table}[hbt]
\caption{The location of critical points $(\Omega_{\text{crit}} , T_{\text{crit}} )$ in the $T-\Omega$ phase portrait of Fig.\ref{phaseTOmega} for different $\mu$.}\label{tab2}
\vspace{1ex}
\begin{tabular}{ccc}
\hline \hline
$\mu$ (MeV) $\qquad \qquad$ & $\Omega_{\text{crit}}$(MeV) $\qquad \qquad$ & $T_{\text{crit}}$(MeV)\\
\hline
$0$ $\qquad \qquad$ &  $707$ $\qquad \qquad$ & $83$ \\
$25$ $\qquad \qquad$ &  $707$ $\qquad \qquad$ & $81$ \\
$70$ $\qquad \qquad$ &  $708$ $\qquad \qquad$ & $72$ \\
\hline \hline
\end{tabular}
\end{table}
%%%%
\par
%%%%
\begin{figure*}[hbt]
\includegraphics[width=5.7cm,height=5cm]{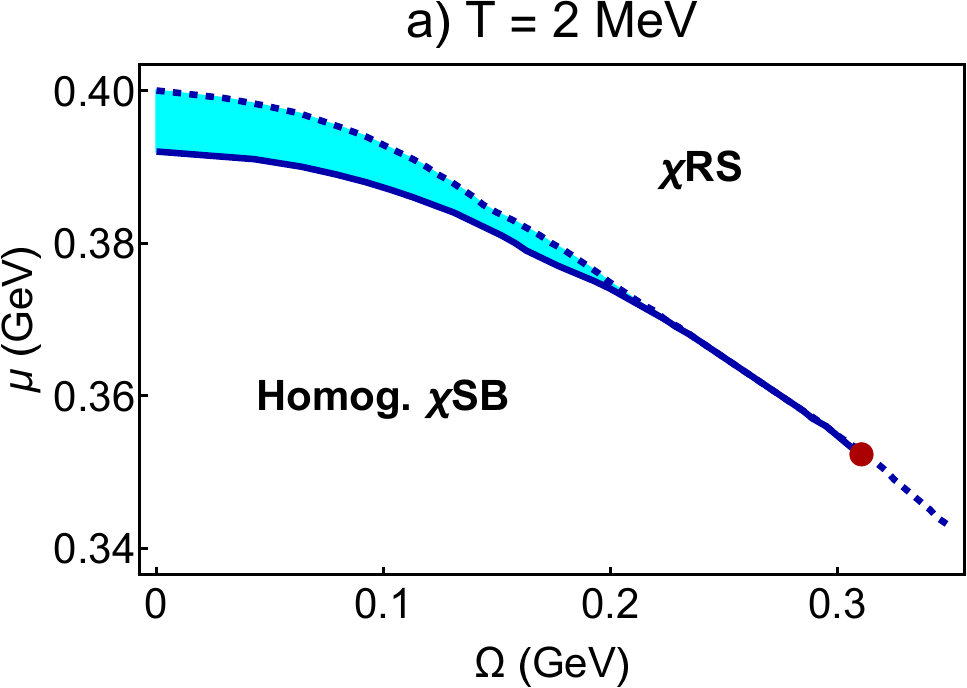}\hspace{0.05cm}
\includegraphics[width=5.7cm,height=5cm]{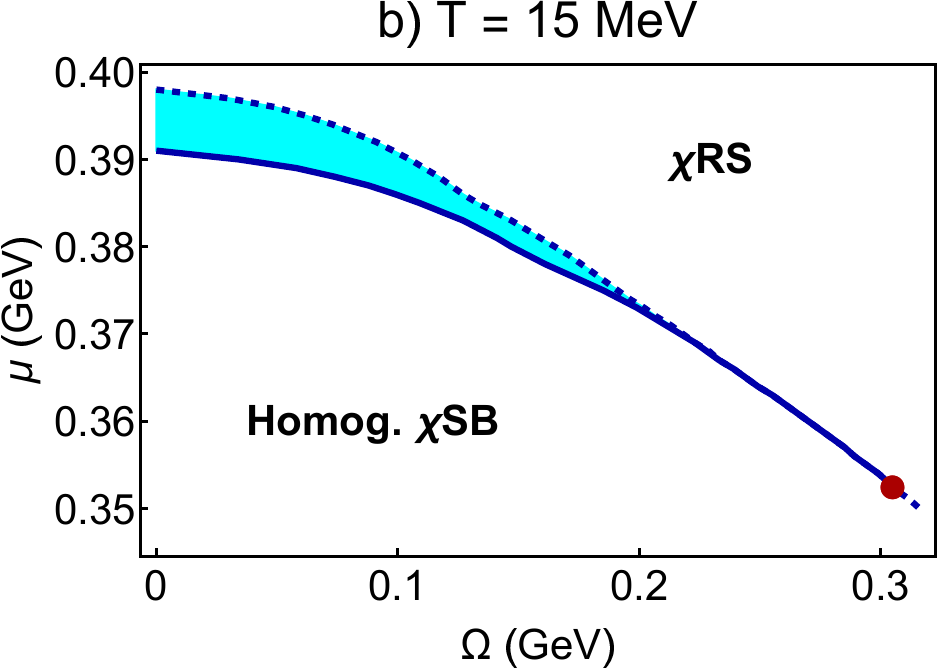} \hspace{0.05 cm}
\includegraphics[width=5.7cm,height=5cm]{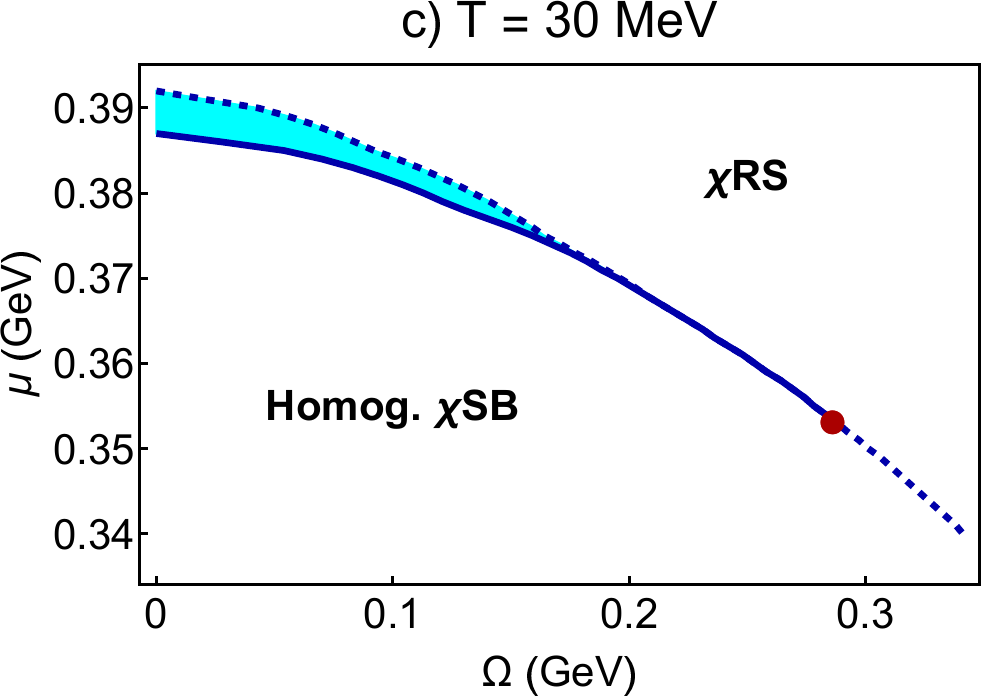}
\vskip 0.1 cm
\includegraphics[width=5.6cm,height=5cm]{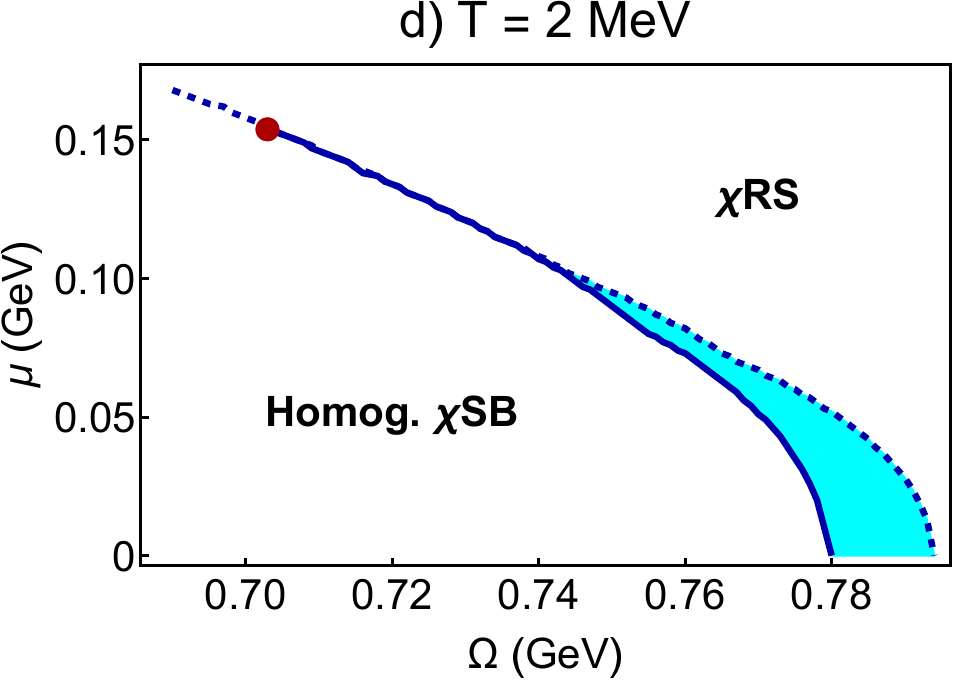} \hspace{0.03 cm}
\includegraphics[width=5.6cm,height=5cm]{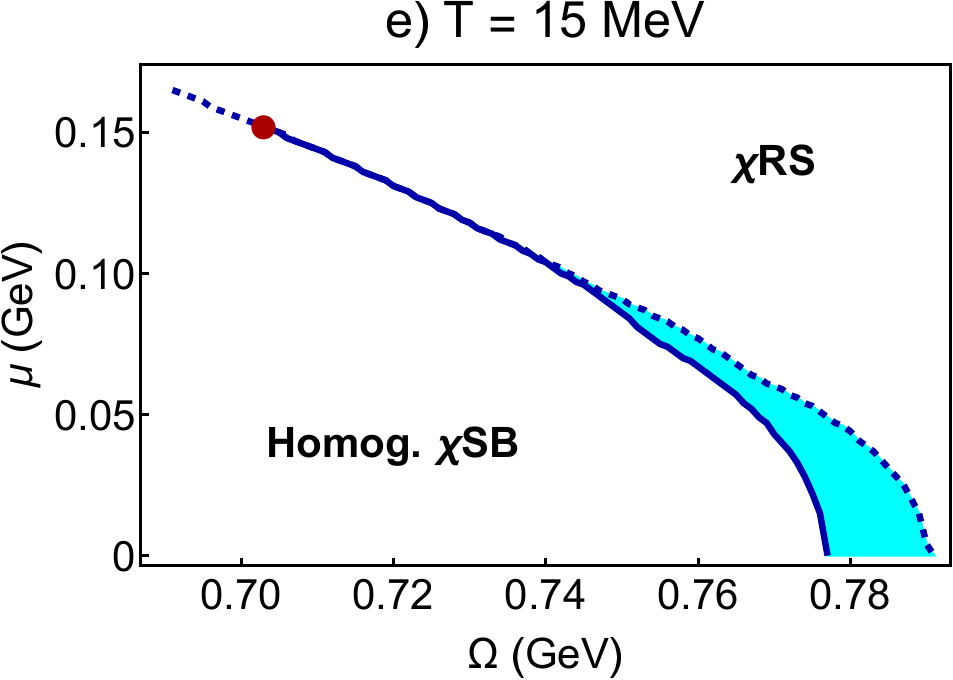} \hspace{0.03 cm}
\includegraphics[width=5.6cm,height=5cm]{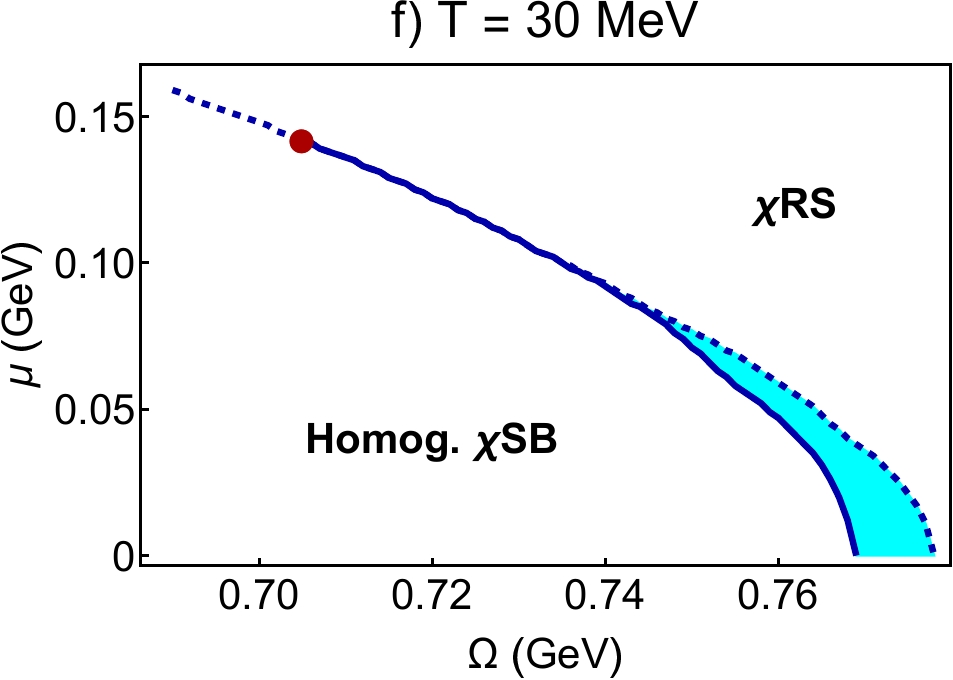}
\caption{(color online). The $\mu - \Omega$ phase diagram of a hot rotating two-flavor NJL model for two cases of small ratio of angular velocity to chemical potential $\frac{\Omega}{\mu} < 1$ (top panel) and large $\frac{\Omega}{\mu} > 1$ (bottom panel). Between these two limits, the DCDW no longer exists and the homogeneous phase goes to the normal phase by a second-order transition. Moreover, the critical point, whose coordinate is calculated by the Ginzburg-Landau analysis, is denoted by the red bullet.}\label{phaseMuOmega}
\end{figure*}
%%%%
%%%%
\begin{table*}[hbt]
\caption{The location of critical points $(\Omega_{1,\text{crit}} , \mu_{1,\text{crit}} )$ (top panel of Fig.\ref{phaseMuOmega}) as well as $(\Omega_{2,\text{crit}} , \mu_{2,\text{crit}} )$ (bottom panel of Fig.\ref{phaseMuOmega})  in the $\mu-\Omega$ phase portrait at different temperatures.}\label{tab3}
\vskip 0.02 cm
\vspace{1ex}
\begin{tabular}{cccccc}
\hline \hline
$T$ (MeV) $ \qquad \qquad$ & $\Omega_{1,\text{crit}}$(MeV) $\qquad \qquad$ & $\mu_{1,\text{crit}}$(MeV) $\qquad \qquad$ & $\Omega_{2,\text{crit}}$(MeV) $\qquad \qquad$ & $\mu_{2,\text{crit}}$(MeV) \\
\hline
$2$ $\qquad \qquad$ &  $311$ $\qquad \qquad$ & $353$  $\qquad \qquad$ &  $703$ $\qquad \qquad$ & $154$\\
$15$ $\qquad \qquad$ &  $305$ $\qquad \qquad$ & $353$  $\qquad \qquad$ &  $703$ $\qquad \qquad$ & $152$ \\
$30$ $\qquad \qquad$ &  $286$ $\qquad \qquad$ & $354$  $\qquad \qquad$ &  $705$ $\qquad \qquad$ & $142$ \\
\hline \hline
\end{tabular}
\end{table*}
%%%%
To complete our understanding of the phase structure of a dual-chiral density wave in a rotating system, we arrive at the final piece. To scrutinize the interplay of chemical potential and the angular velocity at different temperatures, we plot in Fig.\ref{phaseMuOmega}, the $\mu - \Omega$ phase portrait at $T = 2, 15, 30$ \text{MeV}. In the top panel of this figure, we observe the formation of DCDW at intermediate densities. This phase of matter persists at a certain region of phase space with $\frac{\Omega}{\mu}< 1$. In analogy to Fig.\ref{phaseTMu} and Fig.\ref{phaseTOmega}, after a first-order transition from the homogeneous to the DCDW, the latter goes to the chiral restored phase in a second-order transition. In this plot, the red bullet indicates the position of CP whose exact location is given in Table \ref{tab3}. As already demonstrated in Fig.\ref{phaseTMu}, thermal effects lead the DCDW to a lower region of chemical potential and decrease the distance between two phase boundaries. In the $\mu - \Omega$ plane, we observe that by increasing the $T$, the area of DCDW becomes smaller and moves to smaller $\mu$. Moreover, in the bottom panel of Fig.\ref{phaseMuOmega}, the formation of DCDW is demonstrated in an ultrafast rotating quark matter at different temperatures. As it is shown, similar to the case with $\frac{\Omega}{\mu} < 1$, the increase in temperature has a decrease in the inhomogeneous phase. It is worth noting that, according to Table \ref{tab3}, in the $\mu-\Omega$ phase plane there are two critical points. For the region of phase plane with $\frac{\Omega}{\mu}< 1$ ($\frac{\Omega}{\mu}> 1$), by increasing the temperature the critical angular velocity $\Omega_{1,\text{crit}}$ (critical chemical potential $\mu_{2,\text{crit}}$) decreases while the critical chemical potential $\mu_{1,\text{crit}}$ (critical angular velocity $\Omega_{2,\text{crit}}$) is almost constant.
\par
Once the $\mu-\Omega$ phase portrait of rotating quark matter is established, it becomes evident at the intermediate densities (top panel of Fig.\ref{phaseMuOmega}), there exists, in particular, a finite region of the inhomogeneous phase of chiral condensate. Moreover, this region is enclosed by two phase boundaries that meet at the critical point $(\Omega_{1,\text{crit}}, \mu_{1,\text{crit}} )$. As discussed earlier, at moderate densities and above the corresponding critical angular velocity $\Omega_{1,\text{crit}} \approx 10^{24} \, \text{Hz}$\footnote{To express the angular velocity in a unit with inverse of length scale i.e., $\text{fm}^{-1}$ we use $10^{22} \, \text{Hz} \sim 0.02 \, \text{fm}^{-1}$ \cite{Sadooghi:2021upd}. Moreover, to obtain this quantity in units of energy we use $1 \, \text{fm}^{-1} \sim 200 \, \text{MeV}$ \cite{Tabatabaee:2020efb}.}, there is no DCDW. This observation might be of interest to the physics of compact stars as well as heavy-ion collisions. Therefore, in what follows we address the relevance of this point on these physical systems. According to theoretical studies on the uniformly rotating neutron stars (NS), depending on the equation of state as well as the temperature of the star, they might reach an upper angular velocity of $\Omega_{NS} \approx 10^{4} \, \text{Hz}$ \cite{Hashimoto:1994}. Comparing $\Omega_{NS}$ with the critical angular velocity of our model $\Omega_{1,\text{crit}}$, it is immediately evident the former is much smaller than the latter, thus, the effect of rotation might manifest itself by moving the phase boundaries to a smaller region of chemical potential and shrinking the distance between the two phase boundaries. On the other hand, in the heavy-ion collision, the effect of rotation may be more significant. According to studies, in the collision of heavy nuclei, the vorticity may reach even higher values of $\omega \approx 30 \, \text{MeV}$ at particular energies or impact parameters \cite{Jiang:2016woz}. Therefore, the change in the phase boundaries is more significant. At this stage, it is worth noting that, apart from the dependence of $\Omega_{1,\text{crit}}$ on the model parameters, it also has a spatial dependence. The corresponding dependence comes through the inhomogeneity along the radial direction as well as the local nature of thermodynamic quantities such as temperature $T(r) \sqrt{1-r^{2}\Omega^{2}}= T_{0}$ \cite{Tolman:1930zza,Tolman:1930ona} and chemical potential $\mu(r) \sqrt{1-r^{2}\Omega^{2}}= \mu_{0}$ with $\mu_{0}$ is the chemical potential on the axis of rotation \cite{Chernodub:2020qah,Landau:1980}. Therefore, by increasing the radial dependence, the system experiences higher temperatures and larger chemical potential. This, in turn, affects the phase structure in Fig.\ref{phaseMuOmega} and the location of $\Omega_{1,\text{crit}}$. Therefore, it might be possible to reach $\Omega_{1,\text{crit}}$ by moving to a distance away from the axis of rotation.
%%%%%%%%%%%%% 
\section{CONCLUDING REMARKS} \label{sec4}
\setcounter{equation}{0}
%%%%%%%%%%%%%
In this paper, we studied the effect of rigid rotation on the formation of inhomogeneous chiral condensate. To do this, we considered the dual-chiral density wave whose inhomogeneous modulation includes the simultaneous scalar and pseudoscalar condensate, according to \eqref{A12}. In the first part of the paper, after introducing the two-flavor NJL model in a rotating system and using the notations in curved space-time, we calculated the thermodynamic potential of a hot-rotating quark matter at finite density in the mean-field approximation. Using the semibosonized Lagrangian density of the NJL model, we then solved the corresponding Dirac equation. We showed the Dirac wave functions are different once the spatially varying scalar, as well as pseudoscalar condensates, are taken into account. The latter plays the role of spin density in the Lagrangian. The difference between the spin components manifests itself in the dispersion relation of particles. As a consequence of this difference, the Fermi surface of two spin components is deformed compared to the conventional one.
%The solution of Dirac equation shed light, in particular, on the effect of inhomogenous chiral condensate on the structure of spiors as well as the energy dispersion.
\par
In the second part of this paper, we solved the corresponding gap equations numerically for constituent mass and the wave vector, at zero and finite temperature, to find their $\mu$ and $\Omega$ dependence. The detailed analysis of corresponding solutions gives us new insight into the interplay between the chemical potential and the angular velocity on the inhomogeneous phase of quark matter. We studied, in particular, at zero temperature, the $\mu$ and $\Omega$ dependence of $m$ and $q$. It has been shown that at $\Omega=0$ three phases appear. At low chemical potential, the homogeneous phase is favored and persists up to a certain chemical potential $\mu < \mu_{1}$. Then, the inhomogeneous phase appears with a standing wave form at moderate densities $\mu_{1} < \mu < \mu_{2}$. For $\mu > \mu_{2}$ the chiral symmetry is restored and DCDW is no longer favored. This is in accordance with the results obtained earlier \cite{Tatsumi:2004dx}. Once we increase the $\Omega$, the spin of particles becomes parallel to the rotation axis. This alignment of spin leads to the suppression of chiral condensate. Thus we expect the homogeneous phase to terminate at smaller $\mu$ and consequently enter the inhomogeneous phase at lower chemical potential. An interesting interplay of chemical potential and the angular velocity manifests itself in the case, where after some critical $\Omega$ the homogeneous phase goes to the chiral restored phase in a second-order transition. In this region, DCDW is no longer favored. The region of the $\mu - \Omega$ plane with a dual chiral density wave is demonstrated in the corresponding phase diagram at zero temperature (see Fig.\ref{fig3}).
\par
In the finite temperature case, we numerically determined the $T - \mu$ phase portrait for various $\Omega$. It has been shown that at $\Omega = 0$, the thermal effects shrink the inhomogeneous phase, leading DCDW to exist in a limited region of phase space. This result completely agrees with the previous studies on DCDW and has already been discussed in more detail in \cite{Nakano:2004cd}. Apart from the formation of DCDW at finite density, the effect of $\Omega$ on this phase is striking. It turns out, the rotation shifts the DCDW to a lower region of chemical potential. Moreover, this shift is accompanied by an additional reduction in the first-order phase boundary which separates the homogeneous phase from the dual chiral density wave. This is the result of the change in the location of the critical point in the phase portrait. As the results of the Ginzburg-Landau analysis showed, the critical temperature goes to smaller values while the critical chemical potential is almost constant.
\par
As concerned the effect of rotation, we then studied the phase diagram of our model in the $T - \Omega$ plane. It was demonstrated that at $\mu = 0$ and large angular velocity, the dual chiral density wave appears as the favored phase in a region of phase space. By increasing the chemical potential we observed, an analogous effect as in the $T - \mu$ phase diagram, the DCDW starts to move to lower $\Omega$, and the area shrinks as well. Combining the results obtained from the phase structure of quark matter in $T - \mu$ and $T - \Omega$ planes, we arrive at the $\mu - \Omega$ phase portrait of DCDW. We showed, in this case, the region of phase space that supports the formation of the dual chiral density wave whereas in the remaining region of phase space, the homogeneous chiral condensate is separated from the normal phase with a second-order phase boundary.
\par
At this stage, let us also note that, although the main motivation of the current study comes from the experimental results from the HIC program, the results obtained here, may have some astrophysical implications for the compact stars \cite{Carignano:2015kda}. These compact objects may have a quark core, are rotating around their axis, and subjected to the large magnetic field of $eB \approx 10^{12} \text{G}$ \cite{Watts:2016uzu,Thompson:1996pe}. The existence of magnetic fields changes the dynamic of quark matter \cite{Tabatabaee:2020efb} as well as the thermodynamics and properties of DCDW \cite{Frolov:2010wn,Ferrer:2021mpq}. Thus, it is interesting to extend this work to a case with an external magnetic field at finite temperature and density. We leave this subject for our future studies.

%\begin{acknowledgments}
%I would like to thank...
%\end{acknowledgments}

\appendix

\section{Thermodynamic potential of rotational dual chiral density wave } \label{appendixA}
\subsection{Fermion determinant }

In this appendix, we present the calculation of the thermodynamic potential of interacting quark matter in a rotating frame using the two-flavor NJL model. To determine the one loop effective action of the theory, $\Gamma_{\text{eff}}$, we integrate out the fermionic degrees of freedom using the path integral
\begin{eqnarray}\label{Ap1}
\mathcal{Z} \equiv e^{i \Gamma_{\text{eff}}} = \int \mathcal{D} \bar \psi \mathcal{D} \psi  \exp \left( i \int d^{4}x \sqrt{-g} \mathcal{ L} \right),
\end{eqnarray}
with $g \equiv \det g_{\mu \nu}$ and the Lagrangian density of the model is defined in \eqref{A9}. As it turns out, in order to perform the integration over fermion fields, it is customary to bring the path integral into a more appropriate form. Therefore, we introduce the mesonic auxiliary fields $(\sigma , \bm{\pi})$,
\begin{eqnarray}\label{Ap2}
&&1 = \frac{1}{\mathcal{N}} \int \mathcal{D} \vec \pi \exp \left[ -\frac{i}{2G} \int d^{4}x \left( \bm{\pi} + G \bar \psi i \gamma^{5} \bm{\tau} \psi \right)^{2} \right],  \nonumber \\
&&1 = \frac{1}{\mathcal{N}} \int \mathcal{D} \sigma \exp \left[ -\frac{i}{2G} \int d^{4}x \left( \sigma + G \bar \psi \psi \right)^{2} \right],
\end{eqnarray}
where $\mathcal{N}$ is a numerical factor. Plugging the identity \eqref{Ap2} into \eqref{Ap1}, the semibosonized effective action  in the mean field approximation reads
%\begin{widetext}
\begin{eqnarray}\label{Ap3}
e^{i \Gamma_{\text{eff}}[\sigma , \pi]} &=& \int \mathcal{D} \bar \psi \mathcal{D} \psi  \exp \left( - \frac{i}{2G} \int d^{4}x (\sigma^{2} + \bm{\pi}^{2}) \right. \nonumber \\
&& \left. + i \int d^{4}x \bar \psi \left[ \Pi - \sigma - i \gamma^{5} \bm{\tau} \cdot \bm{\pi} + \mu \gamma^{0} \right] \psi \right). \nonumber \\
\end{eqnarray}
%\end{widetext}
Being interested in the DCDW phase, we assume the spatial dependence of scalar and pseudoscalar is given according to the following configurations
\begin{eqnarray}\label{Ap4}
\sigma &\equiv& -G \langle \bar \psi \psi \rangle = m \cos (q z), \quad \pi_{1}=\pi_{2}=0, \nonumber \\ 
\pi^{0} &\equiv& -G \langle \bar \psi i \gamma^{5} \tau_{3} \psi \rangle = m \sin (q z).
\end{eqnarray}
 It is worth noting that nonvanishing condensates in \eqref{Ap4} are identified with the scalar $\sigma$ and the neutral pion $\pi^{0}$ condensates. 
\par
At this stage, after plugging \eqref{Ap4} into \eqref{Ap3} and performing the chiral transformation $\psi \rightarrow e^{- i \gamma^{5}\tau_{3} q z/2} \psi $, the effective action, $\Gamma_{\text{eff}} = \Gamma^{(0)}_{\text{eff}} + \Gamma^{(1)}_{\text{eff}}$ is derived. It is given in terms of the tree-level action
\begin{eqnarray}\label{Ap5}
\Gamma^{(0)}_{\text{eff}} =  -\frac{1}{2G} \int d^{4}x \, m^{2},
\end{eqnarray}
and the one-loop correction
\begin{eqnarray}\label{Ap6}
\Gamma^{(1)}_{\text{eff}} = - i  \ln \det \left(  \Pi - m - \frac{q}{2} \gamma^{5} \tau_{3} \gamma^{3} + \mu \gamma^{0} \right).
\end{eqnarray}
In order to find a closed form for $\Gamma^{(1)}_{\text{eff}}$, one may use the similarity transformation of the matrix in \eqref{Ap6}. The similarity transformed  matrix of the form $\mathbb{U}^{-1}\mathcal{M} \mathbb{U}$ with $\mathbb{U}$ being an invertible matrix, has the same determinant as the original matrix $\mathcal{M}$. A first attempt to simplify \eqref{Ap6} is obtained by choosing the transformation of the form $\mathbb{U} = \mathbb{U}^{-1} = \gamma^{5}$,
\begin{eqnarray}\label{Ap7}
i \Gamma^{(1)}_{\text{eff}} &=& \frac{1}{2} \ln \det \left(  \Pi - m - \frac{q}{2} \gamma^{5} \tau_{3} \gamma^{3} + \mu \gamma^{0} \right) \nonumber \\
&+& \frac{1}{2} \ln \det \left( - \Pi - m + \frac{q}{2} \gamma^{5} \tau_{3} \gamma^{3} - \mu \gamma^{0} \right) \nonumber \\
&=& \frac{1}{2} \Tr \ln \left( \tilde \Pi^{2} + m^{2} - \frac{q^{2}}{4} + \frac{q}{2} \gamma^{5} \tau_{3} \left[ \gamma^{3} , \tilde \Pi \right] \right), \nonumber \\
\end{eqnarray}
with $\tilde \Pi = \Pi + \mu \gamma^{0}$ and $[.,.]$ being the commutator. In \eqref{Ap7}, the trace operator ($\Tr$), includes a trace over color, flavor, spinors, as well as four-dimensional space-time coordinate. It has been shown, in previous studies, that the constituent mass, in particular, depends on the radial distance from the axis of the cylinder. To arrive at the final result in \eqref{Ap7}, we worked in the local density approximation, $\partial_{r} m \ll m^{2} $\cite{Jiang:2016wvv,Sadooghi:2021upd,Mehr:2022tfq,Ebihara:2016fwa}. 
\par
\begin{widetext}
To proceed further, we choose $\mathbb{U} = \tau_{1}$, and following the same procedure leads to the \eqref{Ap7}, we arrive, after some calculation, at
\begin{eqnarray}\label{Ap8}
i \Gamma^{(1)}_{\text{eff}} &=& \frac{1}{4} \ln \det \left[ \left( ( i \partial_{t} +\Omega \hat{J}_{z} + \mu)^{2} - \mathbb{L}_{+} \mathbb{L}_{-}  + \partial_{z}^{2} - m^{2} + \frac{q^{2}}{4} \right)^{2} - q^{2} \left( (i \partial_{t} +\Omega \hat{J}_{z} + \mu)^{2} - \mathbb{L}_{+} \mathbb{L}_{-}   \right) \right],
\end{eqnarray}
%\end{widetext}
with $\mathbb{L}_{\pm} \equiv i \partial_{x} \pm  \partial_{y}$. Noting that all the operators on the right hand side of \eqref{Ap8} commute, thus, we proceed to build the common basis of eigenfunction. The corresponding normalized wave function reads
\begin{eqnarray}\label{Ap9}
\langle t, \bm r | k^{0}, \bm k \rangle &=&  e^{ -i k^{0} t+ i k_{z} z + i \ell \varphi} \, J_{\ell} (k_{\perp} r),
\end{eqnarray}
with $J_{\ell} (x) $ being the Bessel function of the first kind, $\bm r = (r,\varphi,z)$ and $\bm k = ( k_{\perp},\ell, k_{z})$. This eigenfunction satisfies the completeness 
\begin{eqnarray}\label{Ap10}
 \sum_{\ell = - \infty}^{\infty} \int \frac{d^{4}k}{(2 \pi)^{4}} \, \langle t, \bm r | k^{0}, \bm k \rangle \langle k^{0}, \bm k | t^{\prime}, \bm r^{\prime} \rangle  =  \delta (t -t^{\prime}) \, \delta (z - z^{\prime}) \, \delta (\varphi - \varphi^{\prime})  \, \frac{\delta (r - r^{\prime})}{r},
\end{eqnarray}
and  the orthogonality condition
\begin{eqnarray}\label{Ap11}
\int d^{4}x \, \langle k^{0}, \bm k | t, \bm r \rangle \langle t, \bm r |k^{0 \prime}, \bm k^{\prime}\rangle =  2 \pi \, \delta_{\ell,\ell^{\prime}} \, \delta (k^{0} - k^{0 \prime}) \, \delta (k_{z} - k_{z}^{\prime}) \, \frac{\delta (k_{\perp} - k_{\perp}^{\prime})}{k_{\perp}},
\end{eqnarray}
with $d^{4}x = \text{dt} \, \text{r} \text{dr} \, \text{d}\varphi \, \text{dz}$ being the volume element.
\par
To compute the determinant \eqref{Ap8}, in the momentum space, after using $\ln \det A = \Tr \ln A$, we perform the trace over space-time by inserting an appropriate set of coordinate bases. Moreover, using the expansion of unit operator \eqref{Ap10}, the one-loop effective action reads
%\begin{widetext}
\begin{eqnarray}\label{Ap12}
i \Gamma^{(1)}_{\text{eff}} &=& \tr  \int d^{4}x  \langle t, \bm r |  \ln \left(  \Pi - m - \frac{q}{2} \gamma^{5} \tau_{3} \gamma^{3} + \mu \gamma^{0} \right) | t , \bm r\rangle \nonumber \\
&=&  \tr \sum_{\ell} \int \frac{d^{4}k}{(2 \pi)^{4}} \int d^{4}x \,  \langle t, \bm r |k^{0}, \bm k \rangle \langle k^{0}, \bm k|  \ln \left(  \Pi - m - \frac{q}{2} \gamma^{5} \tau_{3} \gamma^{3} + \mu \gamma^{0} \right) | t , \bm r \rangle \nonumber \\
&=&  \frac{N_{f} N_{c} T}{2} \sum_{s = \pm} \sum_{\ell} \int \frac{d^{4}k}{(2 \pi)^{4}} \int d^{4}x \left( J^{2}_{\ell} (k_{\perp} r) + J^{2}_{\ell + 1} (k_{\perp} r) \right) \ln \left[ - (k^{0} + \Omega (\ell + 1/2) + \mu)^{2} + \epsilon^{2}_{s} \right]
\end{eqnarray}
%\end{widetext}
with $\tr$ acting on the flavor, color, and spin degrees of freedom. Here, the energy spectrum of up and down quarks in a system, developing DCDW configuration \eqref{Ap4} is given by
\begin{eqnarray}\label{Ap13}
\varepsilon_{\pm} = \sqrt{k^{2}_{\perp} + \left ( \sqrt{k_{z}^{2} + m^{2}} \pm \frac{q}{2} \right )^{2}}.
\end{eqnarray}
Moreover, noting that the matrix $\Omega \sigma^{12}/2$ reduces to $\pm \Omega/2$ with the help of spin projection operators, $P_{\pm} \equiv \frac{1}{2} \left( 1 + i \gamma^{1} \gamma^{2} \right)$, the one-loop effective action simplified further by inserting the unit operator $P_{+} + P_{-} = 1$ and using $i \gamma^{1} \gamma^{2} P_{\pm} = \pm P_{\pm}$ in the second line of \eqref{Ap12}.
%%%%%%%%%%%%%%%%%%%%%
\subsection{Finite temperature}
%%%%%%%%%%%%%%%%%%%%%
In what follows, we derive the thermodynamic potential, $\Omega_{\text{eff}}$, at finite temperature. Let us define the thermodynamic potential through the relation $ \Omega_{\text{eff}} = - \frac{\log \mathcal{Z}}{\beta V} |_{t \rightarrow - i \tau}$
 %$\Gamma_{\text{eff}} = \frac{i}{V} \int_{0}^{\beta} d \tau \Omega_{\text{eff}}$ 
 where $\beta \equiv T^{-1}$ is the inverse of temperature and $V$ is three-dimensional volume. At this stage, to determine the $\Omega_{\text{eff}}$, one makes the following replacement:
\begin{eqnarray} \label{Ap14}
t \rightarrow -i \tau, \quad k^{0} \rightarrow i \omega_{n}, \quad \int \frac{dk^{0}}{2 \pi} \rightarrow i T \sum_{n= - \infty}^{\infty},
\end{eqnarray}
where the fermionic Matsubara frequency is defined by $\omega_{n} \equiv (2 n + 1) \pi T$. Following the above recipe and performing the sum over Matsubara frequencies, the thermodynamic potential is given by
%\begin{widetext}
\begin{eqnarray}\label{Ap15}
\Omega_{\text{eff}} &=& \frac{1}{V} \int d^{3} x \left [ \frac{m^{2}}{2 G} -  \frac{N_{f} N_{c}}{4 \pi^{2}} \sum_{s = \pm} \sum_{\ell = -\infty}^{\infty} \int_{0}^{\infty} dk_{z}  \int_{0}^{\infty}  dk_{\perp}  k_{\perp} \left( J^{2}_{\ell} (k_{\perp} r) + J^{2}_{\ell + 1} (k_{\perp} r) \right) \times \right. \nonumber \\
&& \left.
    \left \{   \varepsilon_{s}+ T \ln \left( 1 + e^{- \beta ( \varepsilon_{s} + \Omega (\ell + 1/2) + \mu )}  \right)   + T \ln \left( 1 + e^{- \beta ( \varepsilon_{s} - \Omega (\ell + 1/2) - \mu )}  \right) \right \} \right ].
\end{eqnarray}
%\end{widetext}
The vacuum contribution of the above result has an ultraviolet divergence and hence needs to be regularized. Adopting the proper time method, using the integral
\begin{eqnarray}\label{Ap16}
\frac{1}{A^{n}} = \frac{1}{(n - 1)!} \int_{0}^{\infty} d \tau \tau^{n-1} e^{- \tau A},
\end{eqnarray}
we, eventually, arrive at
%\begin{widetext}
\begin{eqnarray}\label{Ap17}
\Omega_{\text{eff}} &=& \frac{1}{V} \int d^{3} x \left [ \frac{m^{2}}{2 G} + \frac{N_{f} N_{c}}{8 \pi^{5/2}} \sum_{s = \pm} \int_{\Lambda^{-2}}^{\infty} \frac{d \tau}{ \tau^{5/2}} \int_{0}^{\infty} dk_{z}  \exp \left[ - \left( \sqrt{k_{z}^{2}+m^{2}} +  \frac{s\, q}{2} \right)^{2}  \tau \right] - \frac{N_{f} N_{c} T}{4 \pi^{2}} \sum_{\lambda = \pm} \sum_{\ell = -\infty}^{\infty} \int_{0}^{\infty} dk_{z} \times     \right. \nonumber \\
&& \left. \int_{0}^{\infty}  dk_{\perp}  k_{\perp}  \left( J^{2}_{\ell} (k_{\perp} r) + J^{2}_{\ell + 1} (k_{\perp} r) \right)  \left \{ \ln \left( 1 + e^{- \beta ( \varepsilon_{s} + \Omega (\ell + 1/2) + \mu )}  \right) + \ln \left( 1 + e^{- \beta ( \varepsilon_{s} - \Omega (\ell + 1/2) - \mu )}  \right) \right \} \right ],
\end{eqnarray}
with $\Lambda$ being the cutoff for the lower bound of proper time integration. Here, to obtain the final expression of $\Omega_{\text{eff}}$, we used the identity \cite{Gradshtein}
\begin{eqnarray}\label{Ap18}
\sum_{\ell=1}^{\infty} J_{\ell}^{2} (x) = \frac{1}{2} \left( 1 - J_{0}^{2} (x) \right), 
\end{eqnarray}
\end{widetext}
then performed the transverse momentum integral. It is worth mentioning that the final result of $\Omega_{\text{eff}}$ is comparable with the one derived in \cite{Nakano:2004cd} in the absence of rotation. 
%%%%%
%%%%%

% The \nocite command causes all entries in a bibliography to be printed out
% whether or not they are actually referenced in the text. This is appropriate
% for the sample file to show the different styles of references, but authors
% most likely will not want to use it.
%\nocite{*}

%\bibliography{apssamp}% Produces the bibliography via BibTeX.
%%%%%%%%%%%%%%%%%%%%

\end{document}